\documentclass[useAMS,usenatbib]{mn2e}
\usepackage{epsfig}
\usepackage{amssymb}
\usepackage{color}

\title[{\rm [C{\sc II}]},  {\rm [C{\sc I}]}  and {\rm CO} emission from cloud formation]{Tracing the formation of molecular clouds via {\rm [C{\sc II}]},  {\rm [C{\sc I}]}  and {\rm CO} emission}
\author[]
{Paul C.~Clark$^{1}$ , Simon C.~O.~Glover$^{2}$, Sarah E.~Ragan$^{1}$ \& Ana Duarte-Cabral$^{1}$ \\
$^{1}$School of Physics and Astronomy, Queen's Buildings, The Parade, Cardiff University, Cardiff, CF24 3AA \\
$^{2}$Zentrum f\"ur Astronomie der Universit\"at Heidelberg, Institut f\"ur Theoretische
Astrophysik, Albert-Ueberle-Str.\ 2, 69120 Heidelberg \\
{\tt email:} paul.clark@astro.cf.ac.uk, glover@uni-heidelberg.de, RaganSE@cardiff.ac.uk, adc@astro.cf.ac.uk}
 
\begin{document}
\maketitle

\begin{abstract}
Our understanding of how molecular clouds form in the interstellar medium (ISM) would be greatly helped if we had a reliable observational tracer of the gas flows responsible for forming the clouds. Fine structure emission from singly ionised and neutral carbon ([CII], [CI]) and rotational line emission from CO are all observed to be associated  with molecular clouds. However, it remains unclear whether any of these tracers can be used to study the inflow of gas onto an assembling cloud, or whether they primarily trace the cloud once it has already assembled. In this paper, we address this issue with the help of high resolution simulations of molecular cloud formation that  include a sophisticated treatment of the chemistry and thermal physics of the ISM. Our simulations demonstrate that both [CI] and CO emission trace gas that is predominantly molecular, with a density $n \sim 500$--1000~cm$^{-3}$, much larger than the density of the inflowing gas. [CII] traces lower density material ($n \sim 100 \: {\rm cm^{-3}}$)  that is mainly atomic at early times. A large fraction of the [CII] emission traces the same structures as the [CI] or CO emission, but some arises in the inflowing gas. Unfortunately, this emission is very faint and will be difficult to detect with current observational facilities, even for clouds situated in regions with an elevated interstellar radiation field.
\end{abstract}


\section{Introduction}
In the Milky Way, and also in other large spiral galaxies, a large fraction of the total gas content is observed to be in the form of massive dense clouds of molecular gas. These molecular clouds are the birthplaces of new stars and star clusters, and hence play a key role in the overall galactic matter cycle. Therefore, if we want to understand how galaxies form stars and regulate their star formation rates, it is important to understand how molecular clouds themselves are formed. 

Efforts to model molecular cloud formation have been going on for decades (see e.g.\ the historical summaries in \citealt{dobbs14} and \citealt{kg16}), and many different mechanisms have been suggested. At the present time, the most favoured models are ones in which clouds form from converging flows of atomic or molecular gas, driven either by stellar feedback processes (supernovae, stellar winds etc.), or by large-scale gravitational instability \citep[e.g.][]{bp99,hp99,hp00,heitsch06,ii08,ji16,pad16,ds17}. However, observational evidence in favour of either of these pictures remains scarce. 

One reason for this is the difficulty involved in finding good observational tracers of the cloud assembly process. Models of cloud formation predict that even if the converging gas flows are initially atomic, molecular hydrogen (H$_{2}$) will form within the flow on a relatively short timescale \citep{bergin04,ii12,clark12}. However, H$_{2}$ is not detectable in emission at the low temperatures ($T < 100$~K) characteristic of gas in the assembling clouds. Therefore, in order to probe the behaviour of the molecular gas during cloud assembly, it is necessary to find some observational proxy for the H$_{2}$ that can be easily mapped.  

The most commonly used observational proxy for H$_{2}$ is carbon monoxide, CO. Unfortunately, although this molecule is an extremely useful probe of the behaviour of H$_{2}$ within molecular clouds that have already formed, it does not appear to be a good tracer of the cloud assembly process. Most clouds that are CO-bright are associated with at least some level of ongoing star formation, suggesting that they have already accumulated enough mass to become gravitationally unstable. Numerical models also predict that the time lag between the formation of detectable quantities of CO and the onset of star formation should be short \citep{hh08,clark12}. Alternatives to CO for tracing the H$_{2}$ content of the gas are therefore of great interest.

One promising possibility is the use of the [CII]$\,$ 158~$\mu$m fine structure line as a tracer of CO-dark molecular gas. This line is the dominant coolant in the neutral interstellar medium \citep{holl91} and hence is one of the brightest emission lines in most galaxies. Furthermore, both observations \citep{pineda13} and simulations \citep{gs16} suggest that a significant fraction of the observed emission originates from H$_{2}$-dominated gas. Nevertheless, the reliability of [CII] emission as a tracer of molecular gas on the scale of individual molecular clouds remains to be established.

Another possibility is neutral atomic carbon, C, which has two fine structure lines at 370~$\mu$m and 609~$\mu$m.  Chemically, C is found in regions intermediate between the lower density, low extinction material traced by [CII] and the higher density, high extinction gas traced by CO. It is therefore more effective than CO at tracing gas at the boundaries of molecular clouds \citep{glover15}, and in some circumstances can be a more reliable tracer of the total molecular mass than CO \citep{offner14,gc16}. It is particularly useful in regions with a high cosmic ray flux, where the transition from C to CO occurs at a much higher density than is the case in local clouds \citep{bisbas15,gc16,bisbas17a}.  However, simulations of [CI] emission from realistic molecular clouds have so far focussed on clouds that have already formed, and hence have been unable to explore whether [CI] is a good tracer of the inflowing gas, or whether, like CO, it primarily traces the cloud once it has already assembled. 

In this paper we produce and analyze synthetic maps of [CII], [CI] and CO emission based on a simulation of the collision of two, initially {\em atomic} clouds. Our goal is to determine what regimes the various tracers can probe -- in terms of density, temperature and molecular (H$_2$) fraction -- and how this varies as we expose the clouds to progressively stronger interstellar radiation fields (ISRFs) and cosmic ray ionisation rates (CRIRs). In particular, we are interested in determining whether any of the tracers are sensitive to the inflowing gas in the clouds, or whether they simply trace the denser molecular cloud formed by the collision.  In this paper, we will focus on the emission from 4 lines: [CII]$\,$ 158~$\mu$m,  CO (1-0) and the first two lines of [CI] : the $^3 P_1 \rightarrow ^3 P_0$ transition at 609~$\mu$m and the $^3 P_2 \rightarrow ^3 P_1$ transition at 307~$\mu$m, which we will refer to as [CI] (1-0) and [CI] (2-1) respectively for convenience. 

The structure of our paper is as follows. in Section~\ref{sec:method}, we outline the numerical method used to carry out the simulations and to produce the synthetic emission maps. In Section~\ref{sec:depend}, we examine how the strength and spatial distribution of the different tracers changes as we increase the strength of the ISRF and the CRIR. In Section~\ref{sec:ppp2ppv}, we discuss how we associate the emission we observe at different velocities with the gas responsible for producing it, and what this tells us about the properties of the gas producing the bulk of the emission. In Section~\ref{sec:vels}, we examine the kinematics of the emission, and in Section~\ref{sec:discuss} we compare our results to those from the GOT-C$^+$ survey of the Milky Way and from previous numerical simulations. We conclude by summarizing our key results in Section~\ref{sec:conc}.

\begin{figure*}
\centerline{ \includegraphics[width=6.8in]{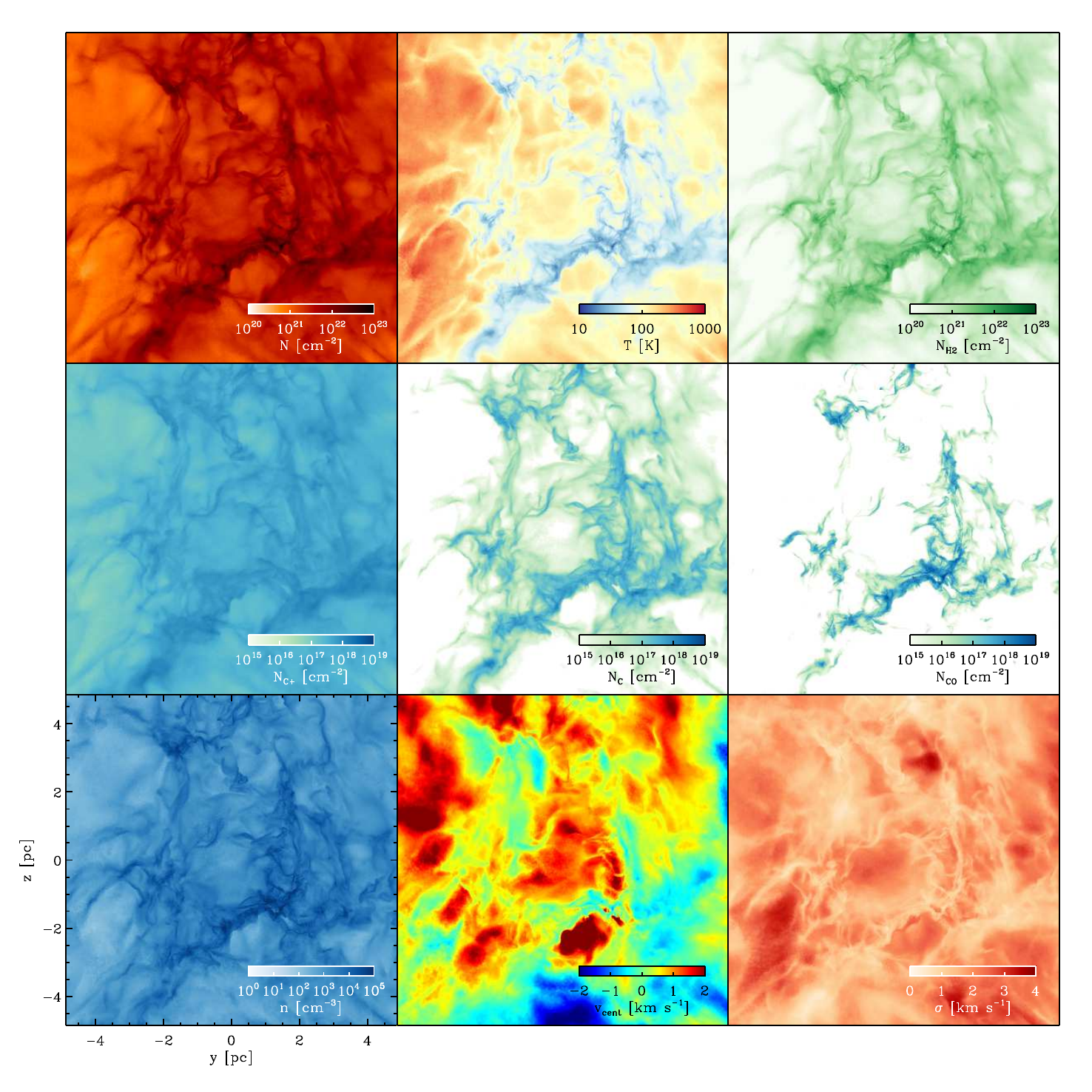} }
\caption{Properties of the gas in the 9pc $\times$ 9pc region in which we perform our radiative transfer analysis in this paper.  We show here only the simulation with G$_0$ = 17 and  CRIR = $3 \times 10^{-16}$ s$^{-1}$, since this simulation has the most striking features.  The bottom left panel shows the mean density along the line-of-sight, weighted by mass.  The bottom-middle and bottom-right panels show the density-weighted velocity centroid and density-weighted velocity dispersion, which are calculated via the expressions $v_{\rm cent} = \sum_i v_x[i]\, \rho[i] / \sum_i \rho[i] $, and $\sigma^2 = \sum_i (v_{\rm cent} - v_x[i])^2 \, \rho[i] / \sum_i \rho[i] $, where $v_x[i]$ is the x-component of velocity for each cell $i$ along the $x$-axis and $\rho[i]$ denotes the cell densities. The top-right panel shows  the column density of hydrogen {\em nuclei}, given by $N = \int \rho / (1.4 \, m_{\rm H})  \, dx$, where $\rho$ is the 3D gas density and $m_{\rm H}$ is the mass of a hydrogen atom. The middle panels show the column density of the three main tracers that we explore in this paper, given by $N = \int x_{\rm spec} \, \rho / (1.4 \, m_{\rm H})  \, dx$, where $x_{\rm spec}$ denotes the abundance of either C$^+$, C or CO relative to the number of hydrogen nuclei. The temperature in the top-middle panel is weighted by column density.}
\label{fig:X10_images}
\end{figure*}

\section{Methodology}
\label{sec:method}
\subsection{Magnetohydrodynamical simulations}
\label{sec:hydro}
The simulations described in this paper were carried out using the {\sc Arepo} moving-mesh code \citep{springel10}. This code solves the equations of fluid flow on an unstructured mesh, defined as the Voronoi tessellation of a set of mesh-generating points that can move freely with the gas flow. {\sc Arepo} is a quasi-Lagrangian code, making it well suited to problems spanning a wide range of spatial scales, such as the cloud-cloud collisions considered here. In addition, mesh cells can easily be refined or de-refined simply by adding or removing mesh-generating points, meaning that {\sc Arepo} shares many of the strengths of modern Eulerian adaptive mesh refinement codes. Our simulations include a magnetic field, and so we use the treatment of ideal magnetohydrodynamics implemented in {\sc arepo} by \citet{pbs11} and \citet{ps13}. This scheme uses the \citet{powell99} method to mitigate the impact of magnetic field divergence errors. 

The version of {\sc Arepo} used for the simulations described here includes a simplified treatment of the chemical and thermal evolution of the interstellar medium (ISM).  We model the chemistry of hydrogen, carbon and oxygen using an updated version of the NL99 network described in \citet{gc12}.This combines the treatment of hydrogen chemistry presented in \citet{gm07a,gm07b} with the simplified network for CO formation and destruction introduced by \citet{nl99}. Our present version of the NL99 network includes several improvements over the version originally presented in \citet{gc12}, such as a more accurate treatment of CO photodissociation. Full details of the current version of the network can be found in Appendix~\ref{app:chem}.\footnote{We have verified that we obtain broadly similar results if we use the revised version of the NL99 network presented in \citet{gong17} in place of the version used here. We briefly discuss these differences in Section \ref{sec:discuss}.}  Radiative heating and cooling of the gas is treated using the detailed atomic and molecular cooling function introduced in \citet{glover10}, updated in \citet{gc12}, and ported to {\sc Arepo} by \citet{smith14}. Both the chemistry and thermal balance are evolved on-the-fly with the hydrodynamics, rather than in a post-processing step.

We use the {\sc treecol} algorithm \citep{cgk12} to compute an approximate $4\pi$ steradian map of the dust extinction and the column densities of H$_{2}$ and CO surrounding each {\sc Arepo} cell. These values are then used to compute the attenuation of the ISRF due to molecular self-shielding and dust absorption, using shielding functions taken from \citet{db96} and \citet{visser09}.

\begin{figure}
\centerline{ \includegraphics[width=3.6in]{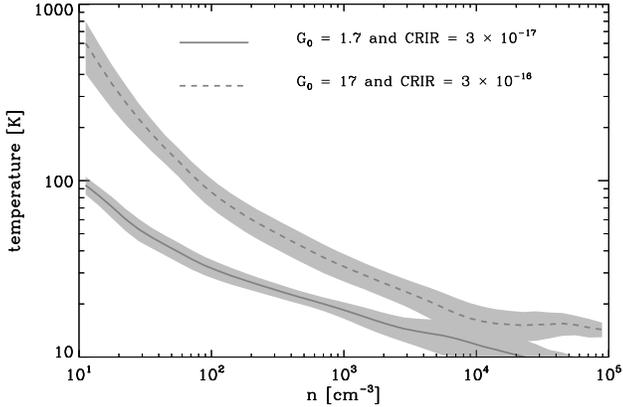} }
\caption{Mean temperature as a function of number density $n$ for our two extreme cases of ISRF and CRIR; we omit the third for clarity. The shaded grey region denotes one standard deviation in the intrinsic scatter in the temperatures at each density.}
\label{fig:rhotemp}
\end{figure}

\subsection{Initial conditions}
\label{sec:ICs}

For our initial conditions, we chose to model the collisions between two atomic clouds, both of which start at a number density of 10 cm$^{-3}$, have a mass of 10$^{4} \rm M_{\odot}$  and a radius of 19 pc. The clouds are given a turbulent velocity field with a velocity dispersion of 1 km s$^{-1}$, which is chosen to follow a standard $P(k) \propto k^{-4}$ scaling law, and in which only solenoidal modes are present. The clouds are set up  to collide head-on, with each cloud given a velocity of 3.75 km s$^{-1}$ towards the other cloud -- and along the x-axis in the domain -- such that the combined collisional velocity is 7.5 km s$^{-1}$. We impose a uniform magnetic field of $B_{\rm x} = 3 \mu$G, such that the collision is occurring along the magnetic field lines.

Our clouds are embedded in a neutral material with number density of 0.1 cm$^{-3}$ in a cubic computational domain of side 190 pc. The centres of the clouds are initially 3 cloud radius apart (57 pc).  The MHD boundaries of the box are periodic, but self-gravity is non-periodic. The initial cell mass is approximately $5 \times 10^{-3} \, \rm M_{\odot}$ -- both in the clouds and the low-density, ambient medium -- such that we have approximately 2 million cells in the `clouds' and 280,000 cells in the surrounding material. We also enforce Jeans refinement, such that the thermal Jeans length is resolved by at least 16 {\sc Arepo} cells at all times. Further, we impose the condition that the volume of neighbouring cells differs by no more than a factor of 8. Our spatial resolution is a function of the local density, but is $\sim 0.1$~pc or better throughout the C and CO-rich portions of the clouds, which is sufficient to yield converged results for the chemical composition and observational properties of the clouds \citep{ds17,gong18,joshi18}.

The above initial set-up is used in 3 simulations in this paper, which are designed to probe the typical environmental conditions under which molecular clouds might form.  In these simulations we vary the strength of the ISRF -- here taken to be a combination of \citet{bl94} for optical and longer wavelengths, and the \citet{dr78} fit in the ultraviolet regime -- from a `solar neighbourhood' value of 1.7 in \citet{habing68} units, to values 3 and 10 times stronger, namely 5.1 and 17. At the same time, we also vary the CRIR of neutral hydrogen from $3 \times 10^{-17}\,\rm s^{-1}$ to $9 \times 10^{-17}\,\rm s^{-1}$ and $3 \times 10^{-16}\, \rm s^{-1}$.  The scaling of the ISRF and CRIR is done together in this study, since both of these parameters depend on the rate of nearby star-forming events.

We consider solar metallicity gas, and adopt values for the initial elemental abundances of carbon, silicon and oxygen given by \citet{sem00}: $x_{\rm C, tot} = 1.4 \times 10^{-4}$, $x_{\rm Si, tot} = 1.5 \times 10^{-5}$ and $x_{\rm O, tot} = 3.2 \times 10^{-4}$, where $x_{i}$ denotes a fractional abundance, by number, relative to the number of hydrogen nuclei. 
As mentioned above, we start our simulations in the cold, neutral medium (CNM), where the initial H$_2$ fraction is zero. However, we initialise our clouds with an H$^+$ abundance of 0.01, to account for the ionisation caused by cosmic rays. In practice, this value is higher than the equilibrium value in each case. However the timescale for recombination is short and so the gas in the clouds reaches the correct ionisation fraction in less than 1~Myr. 

Although the recombination results in some cooling, this is quickly offset by the photoelectric heating in the clouds, such that the initial ionisation fraction and temperature come into equilibrium before the collision is underway. Carbon and oxygen are taken to be their singly-ionised and neutral forms respectively.  Silicon does not play any role in our chemical network and so remains in its singly ionised form throughout the calculations. 

\begin{figure}
\centerline{ \includegraphics[width=3.6in]{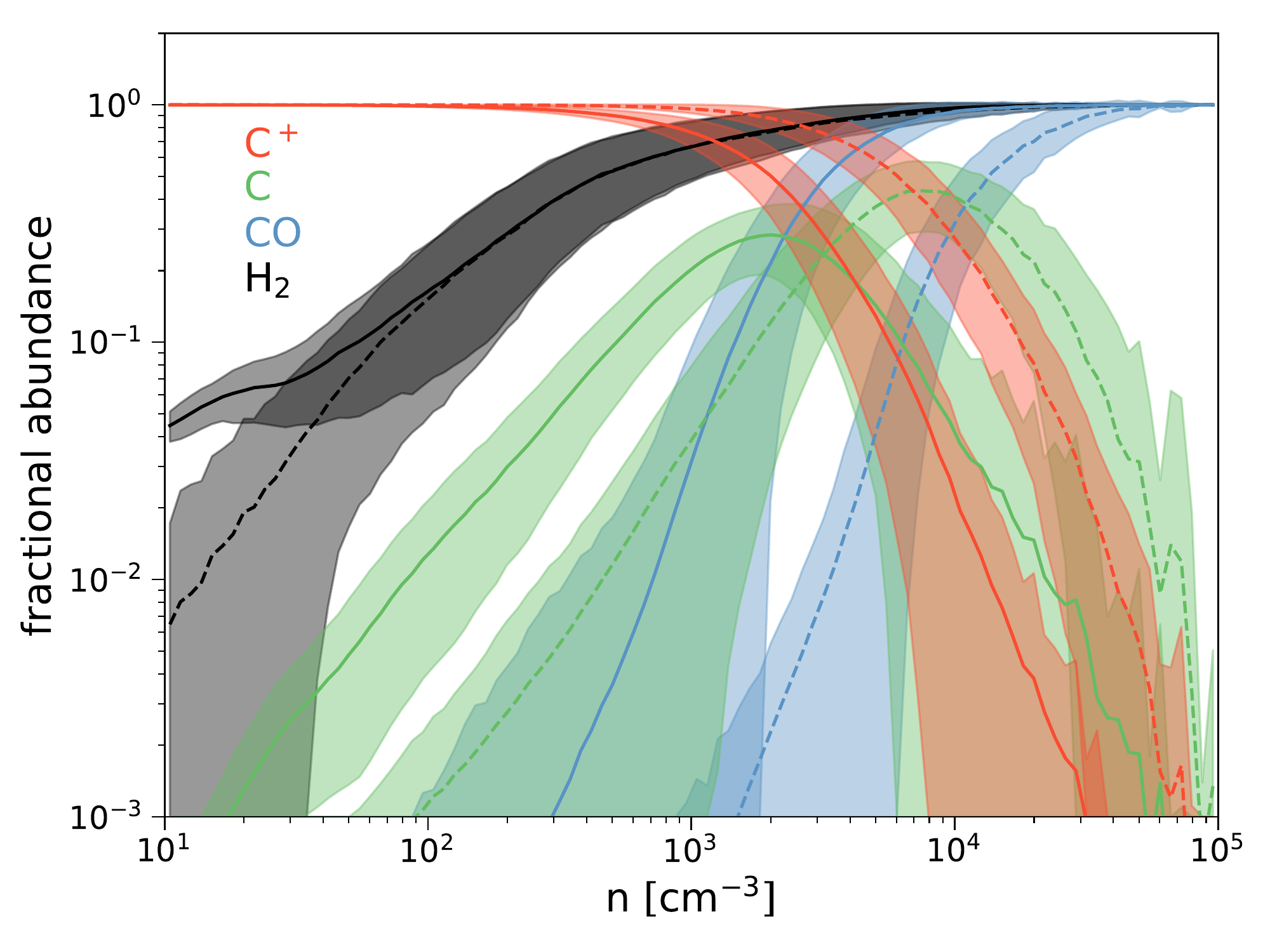}}
\caption{Mean fractional abundances of H$_{2}$,  C$^{+}$, C, CO as a function of number density for our two extreme cases of ISRF and CRIR (again omitting the third for clarity). The shaded regions denote one standard deviation in the abundances. The solid lines show the abundances for the simulation with the solar neighbour hood values of the ISRF and CRIR, and the dashed lines denote the simulation with an ISRF and CRIR ten times higher.  The colours denote the species as shown in the legend.}
\label{fig:rhoxcx}
\end{figure}

\subsection{Radiative transfer post-processing of the simulations}
\label{sec:rt}
We stop our MHD simulations at the point where the first prestellar core starts to undergo gravitational collapse, since we do not include a model for the feedback from young stars.  At this point, we employ the RADMC-3D\footnote{http://www.ita.uni-heidelberg.de/$\sim$dullemond/software/radmc-3d/} code for our radiative transfer (RT) post-processing in this study, making use of the non-LTE line-RT module, using the large velocity gradient approximation  (LVG; \citealt{Sobolev57}) to compute the level populations, as implemented in RADMC-3D by \citet{shetty11a, shetty11b}. We also use the additional `escape probability' option, to limit the length-scale derived by the LVG method to 1 pc. This is useful in regions where the LVG approximation breaks down (e.g.\ in low-density regions where the turbulence is subsonic).   All of the collisional rate data for C$^+$, C, and CO, is taken from the Leiden Atomic and Molecular Database \citep{sch05}.

Our {\sc Arepo} data is interpolated onto a regular, Cartesian grid for the RT post-processing, using routines within {\sc Arepo} to generate the cubes.  Our grids cover a cubic region of size 9.72 pc, with 400 cells per side, giving a spatial resolution of 0.024 pc.  The region is chosen to contain the first star-forming core to form in the simulation, but is also large enough to capture a representative region of the dense, shocked layer that has been formed by the cloud collision. We employ 500 velocity channels in the RT, covering $-5$ to $+5$ km\,s$^{-1}$, giving a channel width of 0.020 km\,s$^{-1}$, sufficient to resolve the line-width in 10 K gas with 10 channels. This resolution is more than sufficient to resolve the high-density cores and filaments within our clouds \citep{penaloza2017}.

In Figure \ref{fig:X10_images} we show the region that will be used for the radiative transfer in this paper. This particular series of images is for the cloud with G$_{0} = 17$ and CRIR $= 3 \times 10^{-16} \: {\rm s^{-1}}$, but exactly the same coordinates are extracted from all our simulations when we perform the radiative transfer. The images in Figure \ref{fig:X10_images} are from the point-of-view of an observer looking {\em along} the collision axis between the two clouds (the $x-$axis), such that we are looking down on the shocked plane, and are made from the 3D cartesian cubes that are imported into RADMC-3D for the radiative transfer.  The analysis presented in this paper will be based around radiative transfer performed along the $x-$axis, such that our velocity channels in the resulting position-position-velocity ($ppv$) cubes are probing the collision between the two clouds.

\section{Dependence on the strength of the ISRF and CRIR}
\label{sec:depend}

\subsection{Physical properties in the clouds}
\label{sec:physprops}

\begin{figure*}
\centerline{ \includegraphics[width=6.8in]{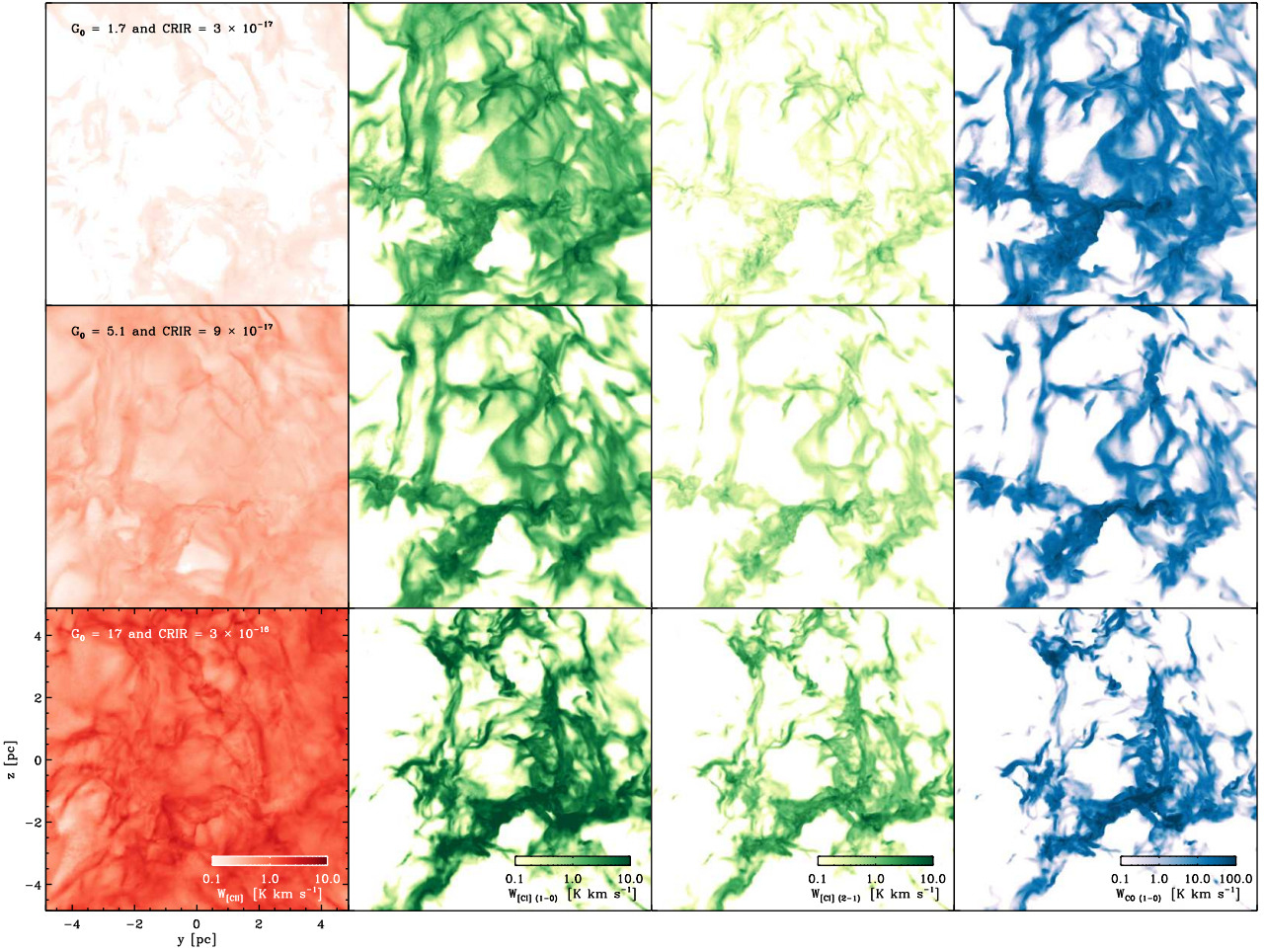} }
\caption{The images show the integrated emission in [CII], [CI] (1-0),  [CI] (2-1), and $^{12}$CO (1-0) from our simulations.}
\label{fig:mom0}
\end{figure*}

Before we look at the line emission from the clouds, we will first examine how the physical properties of the gas depend on the strength of the ISRF and the CRIR. In Figure \ref{fig:rhotemp} we show the variation in gas temperature with number density for two of our runs: the simulation with the standard ISRF and CRIR and the simulation with 10 times higher values for these parameters (denoted by the line styles). The lines follow the mean temperature as a function of density and the grey shaded region denotes one standard deviation about the mean. Overall, we see that the simulation with the higher values for the heating parameters is hotter throughout the entire density regime that we plot here. At low densities, where the shielding from the ISRF is weak, the difference in temperature reflects the amount of photoelectric heating in the clouds. For example,  around a number density of $n = 10\rm \,cm^{-3}$, which is the density we use in our initial conditions, we see that the difference in the gas temperature is around a factor of 8-9. At higher densities, dust shielding starts to become more important, particularly above a density of $10^{3} \, {\rm cm^{-3}}$, and the temperatures start to converge. However, gas in the simulation with high ISRF and CRIR remains hotter than gas in the simulation with low ISRF and CRIR at all densities. In part, this is due to the higher heating rate associated with the higher CRIR -- unlike photoelectric heating, this remains important in high column density gas. In addition, at the highest densities, the difference in the temperature distribution also reflects the growing importance of the thermal coupling between gas and dust\footnote{See also \citet{gold01} for a more detailed discussion of the role played by dust-gas coupling in setting the temperature balance of dense, well-shielded molecular gas.}: the dust is hotter in the simulations with the higher ISRF, since not all of the dust heating comes from the easily-attenuated UV potion of the ISRF. However, at these densities, the gas in the two simulations is also subject to different cooling processes, driven by the differences in the cloud chemistry that we will now discuss.  

In Figure \ref{fig:rhoxcx}, we plot the fractional abundances of H$_2$, C$^+$, C and CO as a function of density for the same two simulations that are represented in Figure \ref{fig:rhotemp}. Once again, the lines denote the mean value and the shaded regions show one standard deviation.   The first feature to note from the plot is that by a number density of $100 \rm \,cm^{-3}$, both of the simulations have about 10 to 20 \% of their hydrogen already in the form of H$_2$, and by $n = 10^3 \rm \,cm^{-3}$ the gas is nearly fully molecular.  

In contrast, the carbon chemistry is significantly more affected by the different levels of the ISRF and the CRIR.  We see that the factor of 10 increase in the ISRF and the CRIR corresponds to a factor of 10 shift in the density at which the transition from C to CO occurs. However, even in the case of a standard ISRF, the majority of the carbon in our clouds is in the form of C$^+$ until at least $2 \times 10^3 \rm \,cm^{-3}$, reiterating the findings of \citet{glover10} and \citet{clark12} that CO resides at densities much higher than we traditionally associate with giant molecular clouds (e.g.\ \citealt{HeyerDame2015}).  This delay in the appearance of the CO in the simulation with the higher values for the ISRF and CRIR also helps to explain the differences in the gas temperature between the runs at densities around $n = 10^4 \rm \,cm^{-3}$, since CO cooling is more efficient than C cooling in this regime and significantly more efficient than C$^+$ cooling.

The transition from C$^+$ to C is also affected by the increase in the ISRF and the CRIR,  but only by a factor of $\sim 4$ in number density. We note that our simplified chemical network may underestimate the C abundance  at the expense of slightly overestimating the C$^+$ abundance \citep{gc12b}. However, this effect is relatively small, and we would not expect this to impact the results presented in the rest of this paper. 

\subsection{Emission from the cloud assembly}
\label{sec:emission}

In this section we look at how the emission from our four lines -- [CII], [CI] (1-0 and 2-1) and CO (1-0) -- varies in the model clouds, and how this relates to the ISFR and CRIR that were adopted in the simulations. We will then relate this to the information on the chemical and thermal properties that we discussed above in Section~\ref{sec:physprops}.

We present in Figure \ref{fig:mom0} the integrated intensity maps from our radiative transfer analysis of the sub-region of the simulations that was described in Section~\ref{sec:rt}. Each row contains the emission maps from one simulation, and the panels in each column show the emission for a different line. Note that the colour scale is stretched over a smaller range of integrated intensities for the fine-structure lines than it is for the CO line. However, in all cases, we show emission down to $\rm 0.1~K\,km\,s^{-1}$. Emission fainter than this is unlikely to be observable with current facilities such as SOFIA, APEX or ALMA (Band 9).

In the right-most column (blue images), we see how the CO emission varies as both the ISFR and CRIR are increased. Overall, we see that the emission becomes brighter and more compact as the heating terms are raised, consistent with the findings of \citet{clark15}. There are three processes responsible for this. The first is that the lower-density gas is hotter in the higher ISRF and CRIR simulation (as shown in Figure \ref{fig:rhotemp} and discussed in Section \ref{sec:physprops}), and pushes on the denser gas, to create sharper, high density features. As these features are well shielded and dense, they are able to harbour CO, as we can see in Figure \ref{fig:X10_images}.  The second phenomenon is simply that the CO-rich gas becomes slightly hotter as the ISRF and CRIR increase (as also discussed in Section \ref{sec:physprops}), which gradually boosts the emission.  The third effect is that the CO in the lower density gas is photo-dissociated, as we have seen in our discussion of Figure \ref{fig:rhoxcx}.

Comparing the integrated emission from the CO and [CI] lines, we see that the maps look very similar in general. However, on closer inspection we see that the [CI] (1-0) is able to trace a larger region than the CO emission, similar to the behaviour discussed in \citet{glover15} and \citet{gc16}, and this becomes more pronounced as we move to higher ISRFs and CRIRs. Given the distributions of C with number density shown in Figure \ref{fig:rhoxcx}, it is clear that this is simply because C is more abundant at lower densities, and thus lower column densities, than CO.   However, the [CI] (2-1) line is less able to trace the lower density gas, owing to its higher critical density. The fact that the second excited level sits at 62.4~K also makes it difficult to excite, since the gas traced by the bulk of the [CI] emission remains relatively cold, as we will discuss further below.  This also explains why we see such an increase in the [CI] (2-1) emission as the ISRF and CRIR are raised, since the line is more easily excited as the gas temperature rises above 20~K.  

While the CO and [CI] emission maps look very similar to one another, tracing roughly the same density structures, the maps of the [CII] emission look very different. Many of the density features that are present in the CO and [CI] emission maps appear as faint outlines in the [CII] maps, consistent with a picture in which C$^+$ surrounds the colder, denser gas.  However, we can also identify structures in the [CII] maps that are totally absent in the CO and [CI] maps, which are associated with features in the column density maps in Figure \ref{fig:X10_images}, demonstrating that the emission from [CII] can trace a different regime to the other lines. 

Although there are clear differences in the [CI] and CO emission from the different simulations, the most striking feature of Figure \ref{fig:mom0} is the large variation in the [CII] integrated intensity as we increase the strength of the ISRF and the CRIR from their fiducial values to values 10 times higher. For the fiducial values, the peak integrated intensity from [CII] is around $\rm 0.23~K\,km\,s^{-1}$ and the peak brightness temperature is $\sim 0.33$~K.  This is marginally detectable with the upGREAT instrument on SOFIA (\citealt{upgreat}, \citealt{gold18}, N.~Schneider, private communication), but only if one focuses on detecting the emission at one or a few locations; mapping the cloud at this level of sensitivity would require a prohibitively large amount of observing time. However, in the runs with high ISRFs and CRIRs, the [CII] emission is significantly brighter, with the peak integrated intensities and brightness temperatures rising to $0.64 \rm~K\,km\,s^{-1}$ and $0.71$~K, respectively, for the run with three times stronger ISRF and CRIR, and 2.55$\rm~K\,km\,s^{-1}$ and 2.66~K for the run with the ISRF and CRIR at 10 times their fiducial values. An important conclusion is therefore that it will be much easier to detect and map the widespread [CII] emission in clouds exposed to radiation field strengths stronger than the local value. We also see that in the cases where the emission is strong enough to be detectable, it is also relatively uniform across the map, providing information on region of both low and high column density. The reason for this uniformity has its origins in the fact that the [CII] emission is powered primarily by photoelectric heating. At low column densities, the total amount of heating occurring along a sight-line varies with the column density -- more gas means more heating, which in turn means more [CII] emission. However, this breaks down at higher column densities, as the gas starts to become shielded against the photoelectric heating; now adding extra column does not strongly affect the total energy radiated by [CII], and hence does not strongly affect the [CII] integrated intensity. 

Given that C$^+$ is relatively abundant right up to its critical density of around 5000 $\rm \,cm^{-3}$ (and beyond) in all our simulations, the differences in the integrated intensities between the three simulations cannot be due to chemical evolution alone. Indeed, the C$^+$ abundance varies by only a factor of 5 at the critical density (and one would expect the emission to become strong at densities below this), while the differences in the integrated intensities are over an order of magnitude.  Rather, the sensitivity of the [CII] line-strength to the heating rate is also due to the temperature of the gas. The energy separation of the upper and lower fine structure levels is $E_{10}/k_{\rm B} = 91.2\,$K and so dense gas, which is typically cold, has trouble populating the level. Since the collisional excitation rate is proportional to $e^{-91.2/T_{\rm kin}}$, where $T_{\rm kin}$ is the kinetic temperature of the gas, slight differences in temperature can have a big effect on the overall emission. For example, at a number density of 1000 $\rm cm^{-3}$, the population of the excited state in our high ISRF and CRIR cloud is a factor of $e^{-91.2/30} / e^{-91.2/20} = 4.6$ larger than in the low ISRF and CRIR cloud, using the approximate temperatures from Figure \ref{fig:rhotemp}. At a number density of 100 $\rm cm^{-3}$, this difference increases to a factor of $e^{-91.2/90} / e^{-91.2/30} = 7.6$. The fact that we see an order of magnitude difference in the emission suggests that most of the [CII] emission that we observe is coming from lower densities, where the difference in temperatures is greater, and hence that most of the emission is sub-thermal. This motivates us to examine the origin of the emission in more detail, which we do in the next section.

\section{Deconstructing the emission: from PPV to PPP}
\label{sec:ppp2ppv}
To help us further explore the origin of the emission in our simulations, we convert our 3D, $x-y-z$ cubes of number density and temperature into $z-y-v_x$ cubes --  the same form as the emission cubes that come out of our radiative transfer post-processing. During this conversion, the density, temperature and abundance cubes are binned into the same velocity bins that represent our ``channels'' in the $ppv$ cubes. Note that while it is possible for density and temperature to be single-valued quantities in the velocity field, the emission from a given point will be spread out in velocity space due to thermal broadening and optical depth effects, which we will refer to here as ``channel blending''. In addition, given that our channels have a finite width, it is often the case that gas at two quite different spatially separated locations can contribute to the same velocity channel. In such a case, we use density weighting in all quantities when producing the final ppv cubes for temperature abundance (density is simply averaged, which amounts to mass weighting, since the cells in the RADMC-3D input cubes have equal volume). As such, there is no exact correlation between the density, temperature, and abundance fields and the resulting  $ppv$ emission cubes. Nevertheless, this conversion from $ppp$ to $ppv$ allows us to pick out trends in the data.

\begin{figure*}

\centerline{
	\includegraphics[width=2.3in]{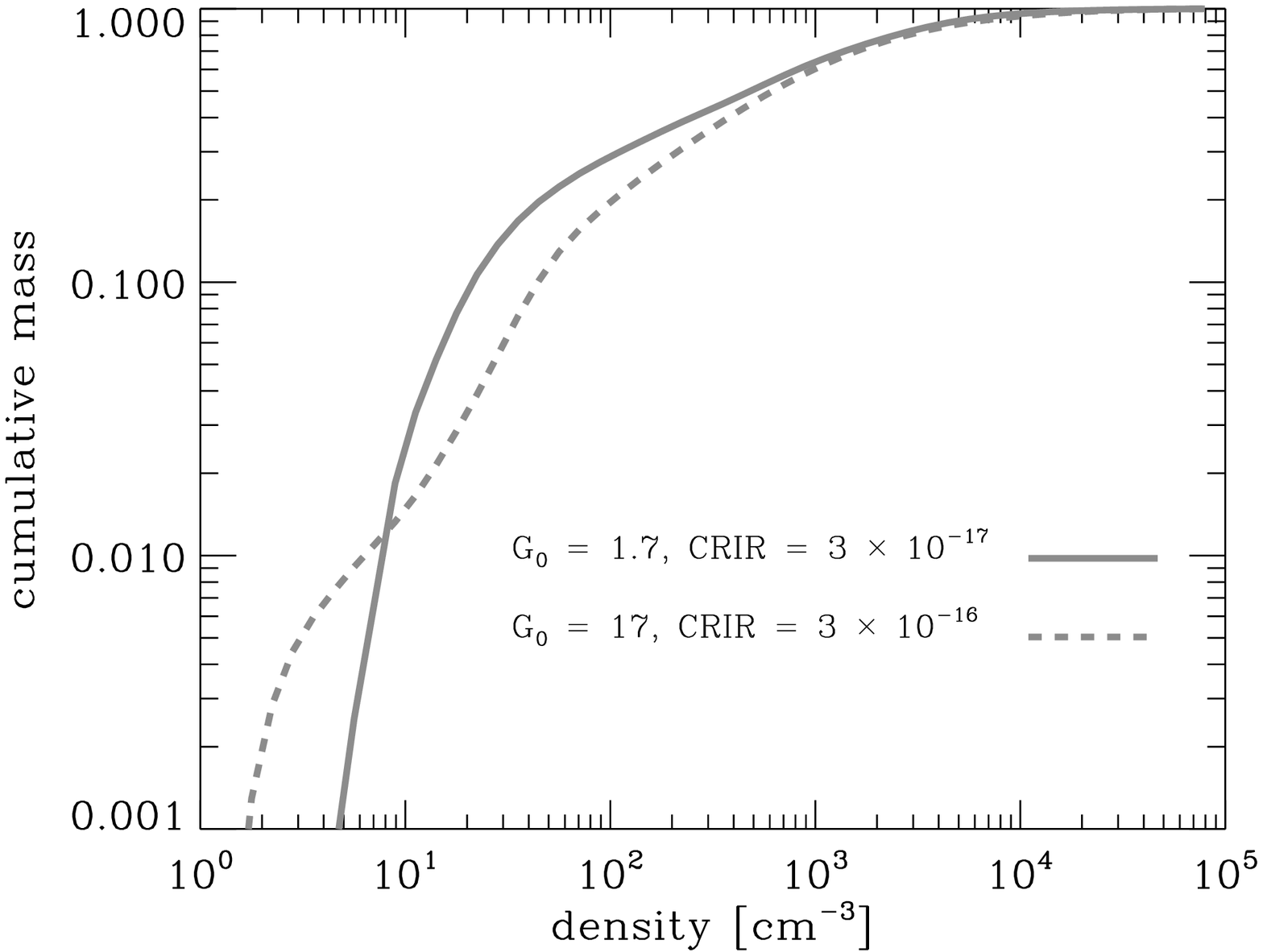} 
	\includegraphics[width=2.3in]{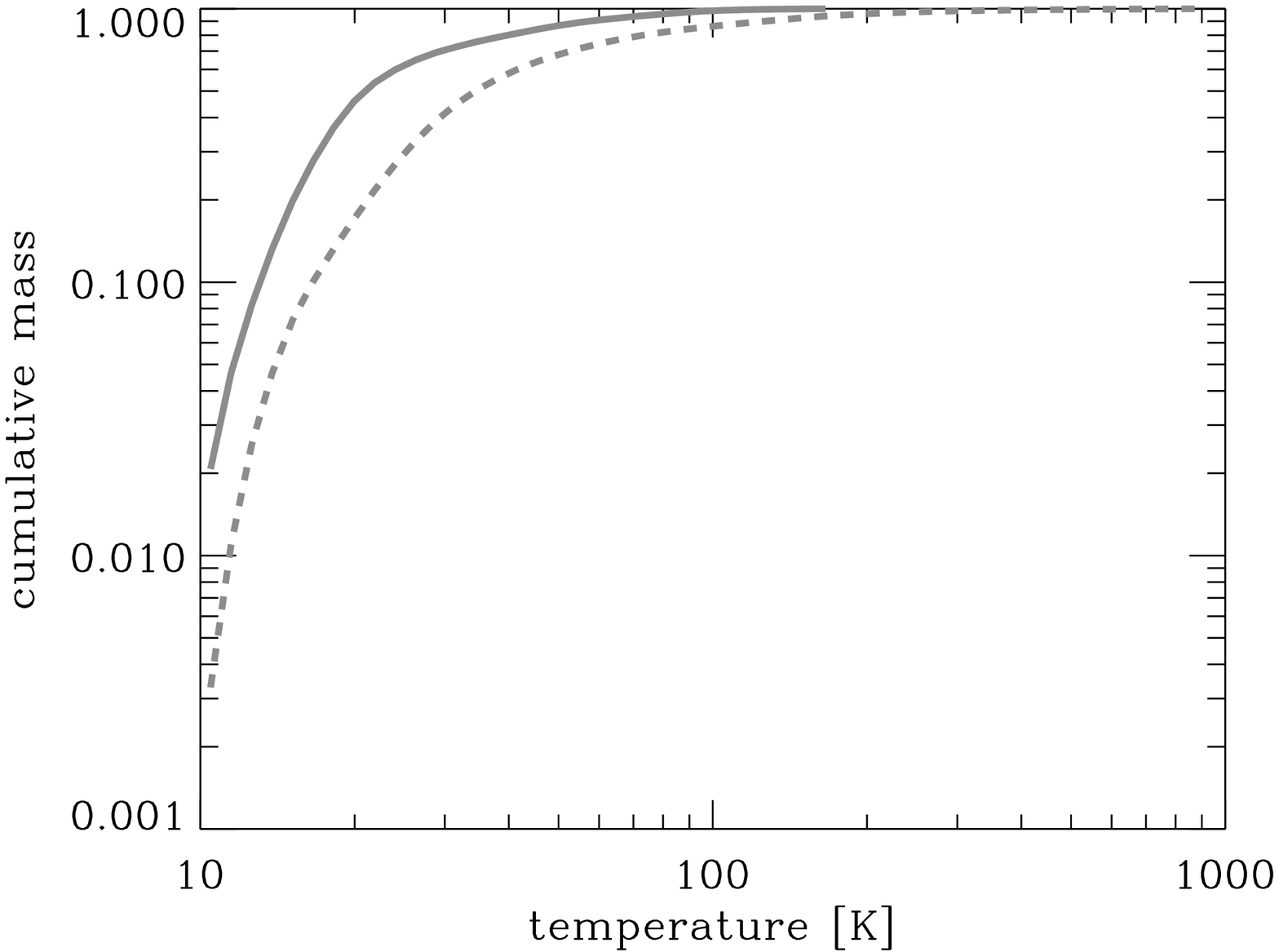}
	\includegraphics[width=2.3in]{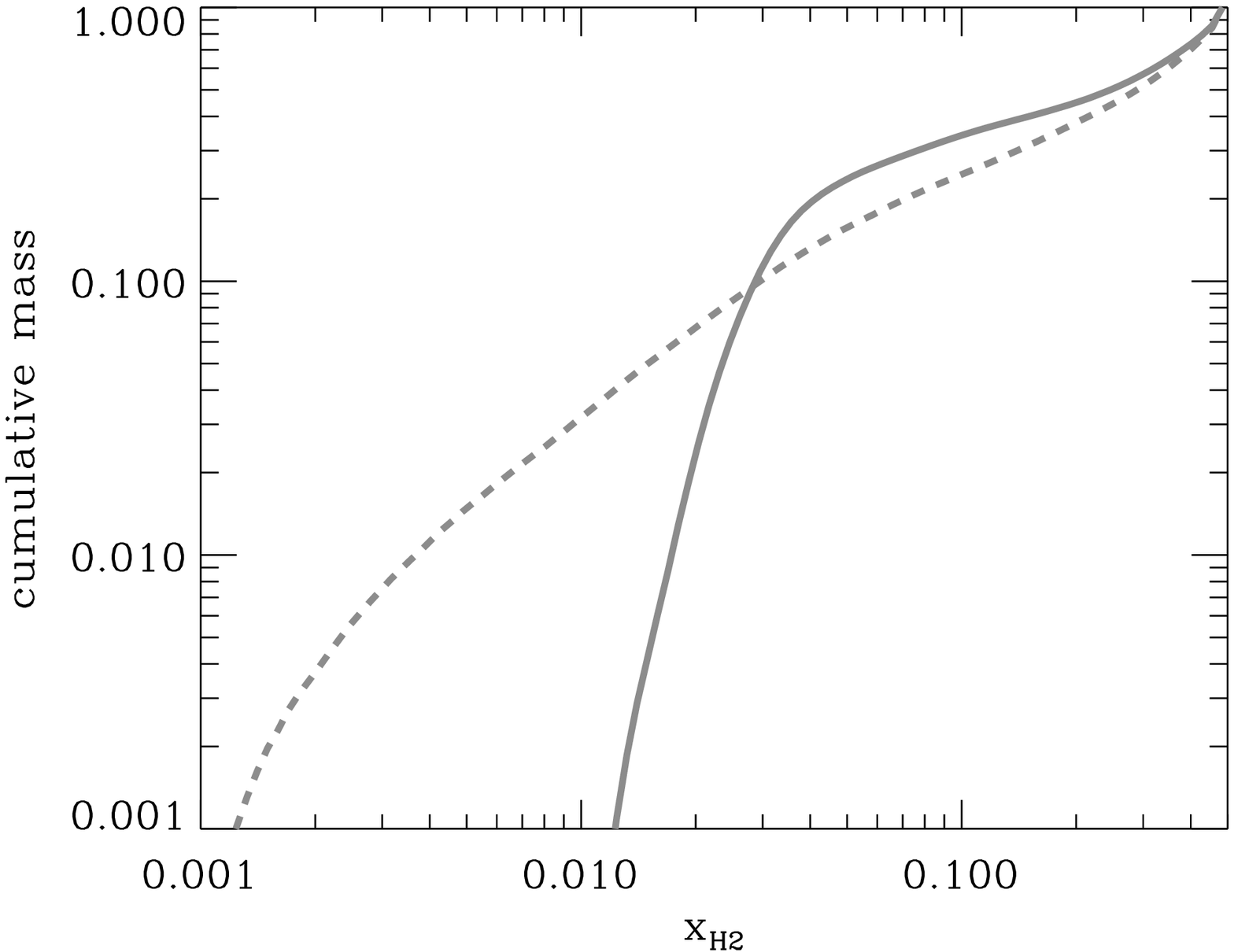}
}
\centerline{
	\includegraphics[width=2.3in]{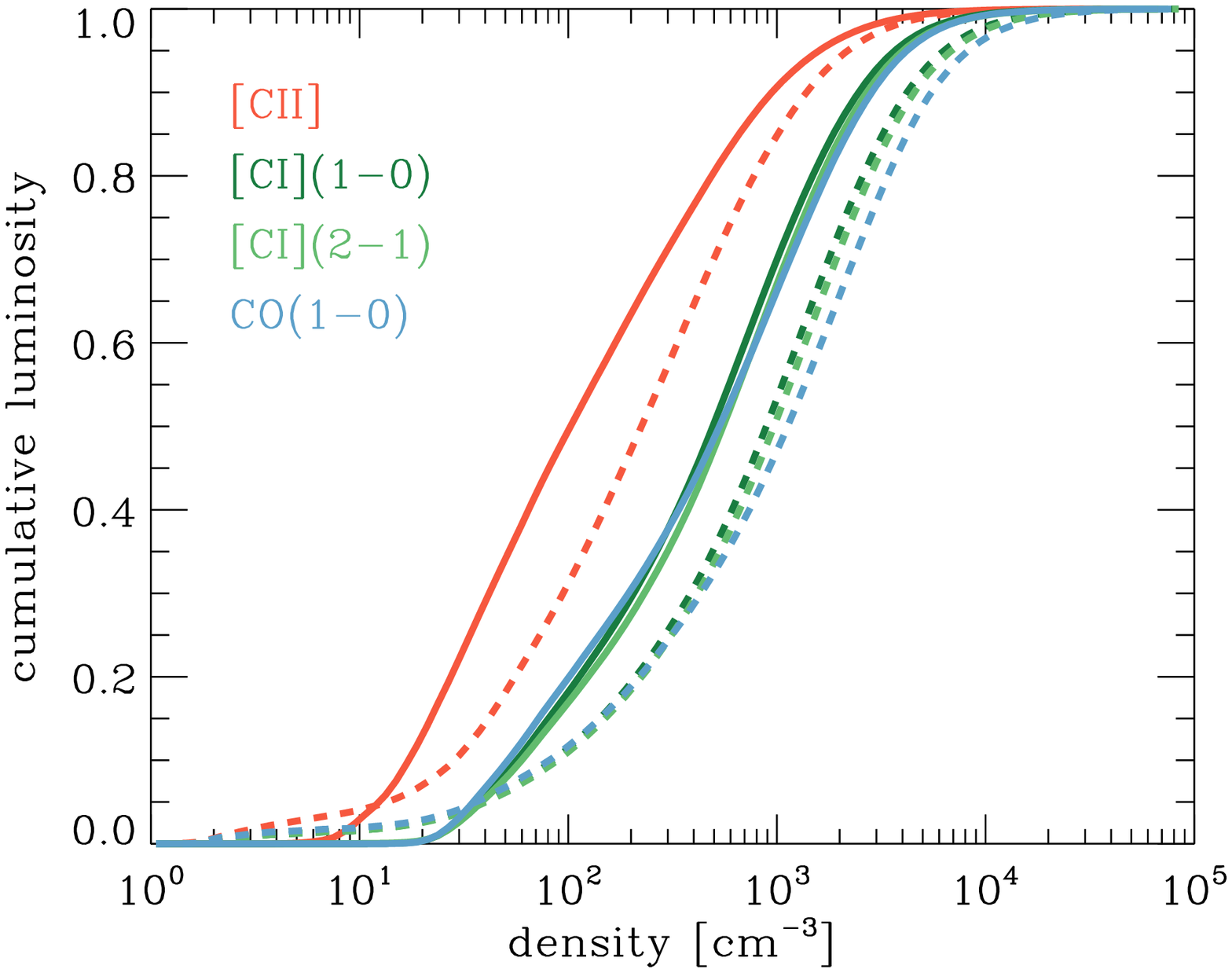} 
	\includegraphics[width=2.3in]{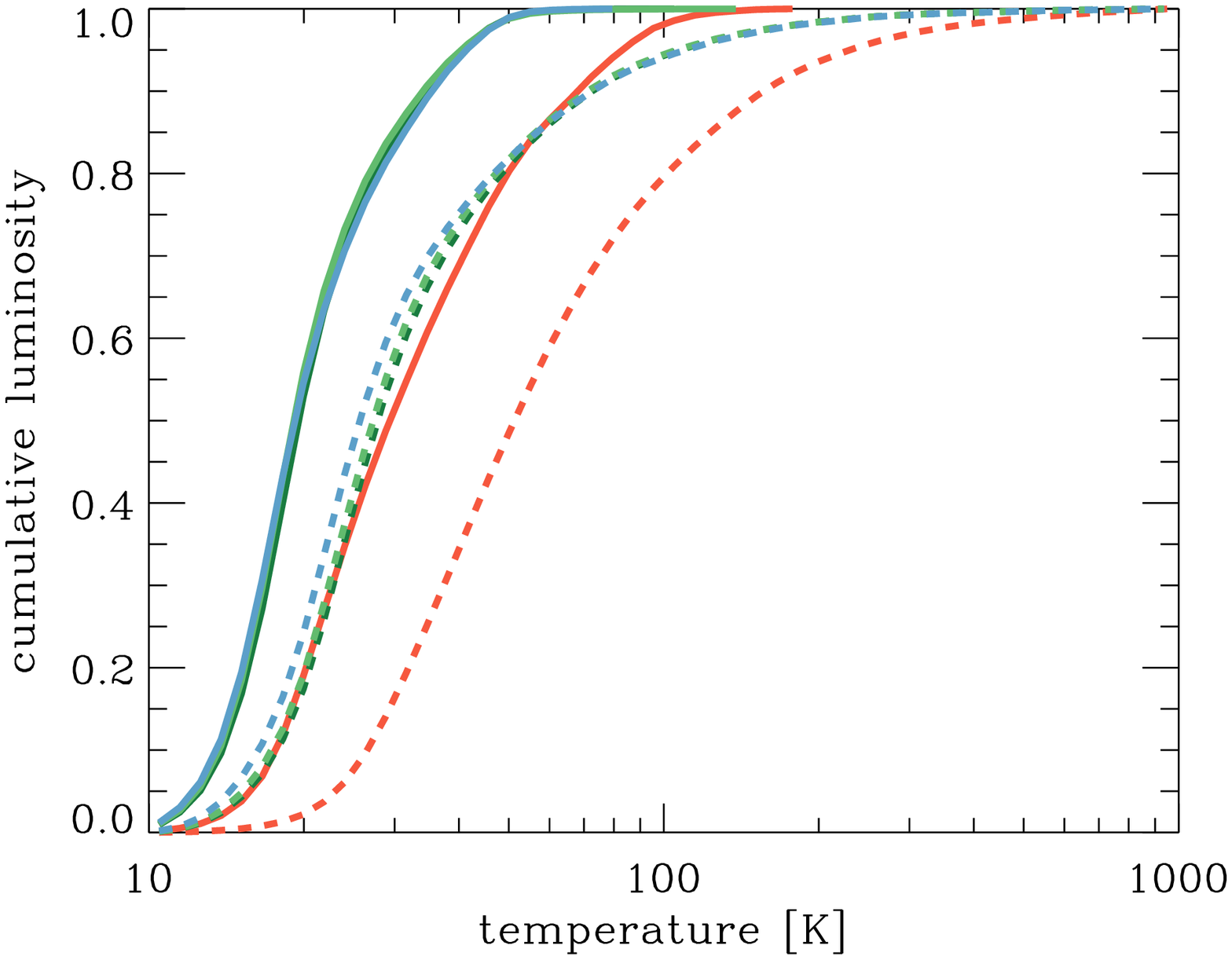}
	\includegraphics[width=2.3in]{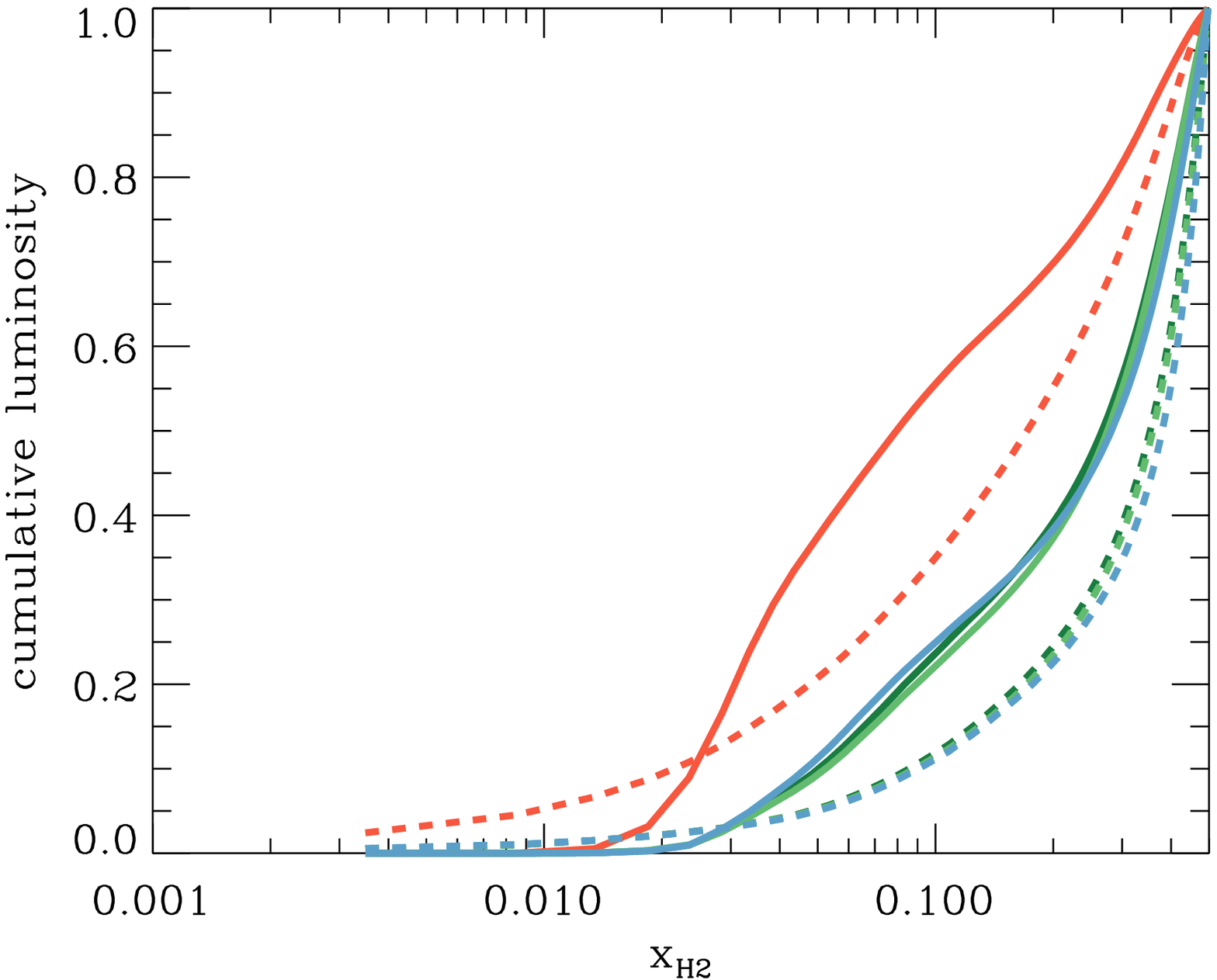}
}
\caption{{\it Top row}: cumulative mass as a function of density, temperature, and atomic hydrogen fraction in our two extreme simulations. As in Figures \ref{fig:rhotemp} and \ref{fig:rhoxcx}, solid lines denote the simulation with G$_0$ = 1.7 and CRIR = $3 \times 10^{-17} \,\rm s^{-1}$, while the dashed lines represent the simulation with values of G$_{0}$ and CRIR that are ten times larger.
{\it Bottom row}: as in the top row, but showing the cumulative luminosity of our different observational tracers. As in previous plots, 
red indicates [CII], dark and light green denote [CI] (1-0) and [CI] (2-1) respectively, and blue represents the CO (1-0) emission.}
\label{fig:cumu}
\end{figure*}

The bottom row of Figure \ref{fig:cumu} shows the cumulative emission of our 4 lines as a function of the 3D properties of the gas. Focusing first on the CO, we see that 50\% of the CO emission traces gas with number densities of around 500 cm$^{-3}$ and higher for the fiducial run, and only 20\% of the emission traces gas residing at number densities below 100 cm$^{-3}$.  With an ISRF and CRIR ten times higher than the fiducial value, we see a slight shift in the cumulative lines -- roughly a factor of 2 in density -- with 50\% of the emission tracing gas above 1000 cm$^{-3}$ and only 10\% tracing gas below 100 cm$^{-3}$.  This confirms that the CO in our simulations is probing the density regime inferred from the observational survey data \citep{RomanDuval2010}.  However, we should stress here that even in the case of the higher ISRF and CRIR, the bulk of the CO emission traces gas sitting below 2200 cm$^{-3}$, the critical density of the (1-0) transition.  Given the high optical depth of CO (1-0), it is likely that this is just the effect of the critical density being lowered due to photon trapping. Although some of the low density CO emission may be sub-thermal, with an excitation temperature lower than the local kinetic temperature, it is likely that the channel blending mentioned above is also playing a role here.\footnote{It would be possible to calculate how much emission is truly sub-thermal by comparing the result here with a series of radiative transfer runs that have had the CO abundance in the cells artificially set to zero below some variable density threshold. However, such a computationally expensive approach is beyond the scope of this study, which is simply to first report which regimes are being {\em} traced by the emission lines.}  If we compare these densities to the plot showing the cumulative mass-fraction as a function of density, we see that the CO is also tracing the bulk of the mass in the region covered by the $ppv$ cubes.

Emission from neutral carbon follows the CO distribution almost exactly, even though its critical density is somewhat lower in the case of the (1-0) line ($\sim 500$ cm$^{-3}$). The explanation for this is given in Figure \ref{fig:rhoxcx}: the abundance of neutral carbon rises sharply with density between 100 cm$^{-3}$ and 1000 cm$^{-3}$, and so below the critical density of the (1-0) line, there is little neutral atomic carbon available to emit \citep[c.f.][who find very similar results]{glover15}.

The emission from the [CII] does appear to have a different origin than the CO and [CI] emission;  roughly 50\% of the emission traces densities around 100 cm$^{-3}$, again with a factor of 2 shift in density between the different ISRF and CRIR values.  However only a few percent of the [CII] emission traces densities similar to those of our original atomic clouds (10 $\rm cm^{-3}$).  This suggests that while [CII] is tracing a different regime than CO, the majority of the emission traces some intermediate phase in the creation of dense (molecular) gas, rather than the low density atomic flows from which the molecular cloud is assembled. We will return to the discussion of these atomic flows in  Section \ref{sec:vels}.

The second panel in the bottom row of Figure \ref{fig:cumu} shows cumulative emission as a function of the gas (kinetic) temperature.  Once again, we see that the CO and [CI] trace similar regimes, with most of the emission probing gas below $\sim 20$ K, and with a factor of less than 2 shift between the different ISRFs and CRIRs.   This tight distribution in the temperatures is a consequence of the temperature-density distributions in Figure \ref{fig:rhotemp} and hence ultimately of the fact that CO and C only become abundant in gas that is effectively shielded from the interstellar radiation field. Because of this shielding, even relatively large changes in the strength of the ambient radiation field yield only small changes in the temperature of the CO-bright gas \citep[see also][who find a similar result using simpler cloud models but explore a wider range of environmental conditions.]{penaloza2017} 

More interesting are the temperatures associated with the [CII] emission. We see that, regardless of the values of the ISRF and CRIR, at least 80\% of the emission traces gas with temperatures below 90 K, the temperature corresponding to the energy difference between the two fine structure levels. In addition, we see that very little of the emission arises from gas with temperatures below 20 K.  This behaviour is a result of the temperature-density structure of the ISM, as shown in Figure \ref{fig:rhotemp}.  Gas warmer than 90K tends to also be low density, and since the emission scales with density as $n^2$ in the sub-thermal regime, the [CII] emission from this warm gas is weak. In contrast, at densities close to or above $n_{\rm crit}$, the gas is cool, with $T \sim 30$~K, and thus struggles to excite the [CII] transition.  As a result, the bulk of the [CII] emission arising from the diffuse ISM comes from a narrow range in density and temperature, imposed by the thermal balance between [CII] cooling and photoelectric heating. 

The right-hand panel in the bottom row in Figure \ref{fig:cumu} shows the cumulative emission as a function of the H$_2$ abundance, $x_{\rm H_{2}}$, and thus reveals which tracers are tracing atomic, versus molecular gas. Looking first at the CO and [CI], we see that in the solar neighbourhood simulation, roughly 55\% of the emission is probing gas with $x_{\rm H_2} \geq 0.3$.  These emission lines are therefore tracing gas which is predominately molecular in composition ($x_{\rm H_2} = 0.5$ denotes gas in which all the hydrogen is in the form of H$_2$).  In the case of the higher ISRF and CRIR simulation, we see that the CO and [CI] now trace gas that has a higher $x_{\rm H_2}$, and is thus more molecular in nature.  This is a result of the photo-chemistry; when we look back to Figure \ref{fig:rhoxcx}, we see that CO and [CI] only become abundant above 1000 cm$^{-3}$, at which point the gas has a high H$_2$ fraction.  

In contrast, we see that the majority of the [CII] emission arises in gas that is mainly atomic in composition. Since the C$^+ \rightarrow$ C transition is delayed until higher densities at higher ISRFs and CRIRs, we see that the [CII] traces more H$_2$-rich gas in these simulations.  However, even in this case, 50\% of the emission is coming from gas with $x_{\rm H_2} \geq 0.2$ and is thus still predominately atomic in composition. 

We can therefore conclude that [CII] traces gas that has started the transition from an atomic to a molecular state, while CO and [CI] trace gas that has nearly finished this transition.  Although [CII] is probing mainly atomic gas, the densities at which the emission becomes detectable is relatively high, at around a few 100 cm$^{-3}$, and due to the thermal balance in the ISM, the gas is generally below 90K.  As such, it appears from our analysis that [CII] is a good tracer of the post-shock structure that directly precedes the formation of molecular clouds.  

\begin{figure}
\includegraphics[width=3.2in]{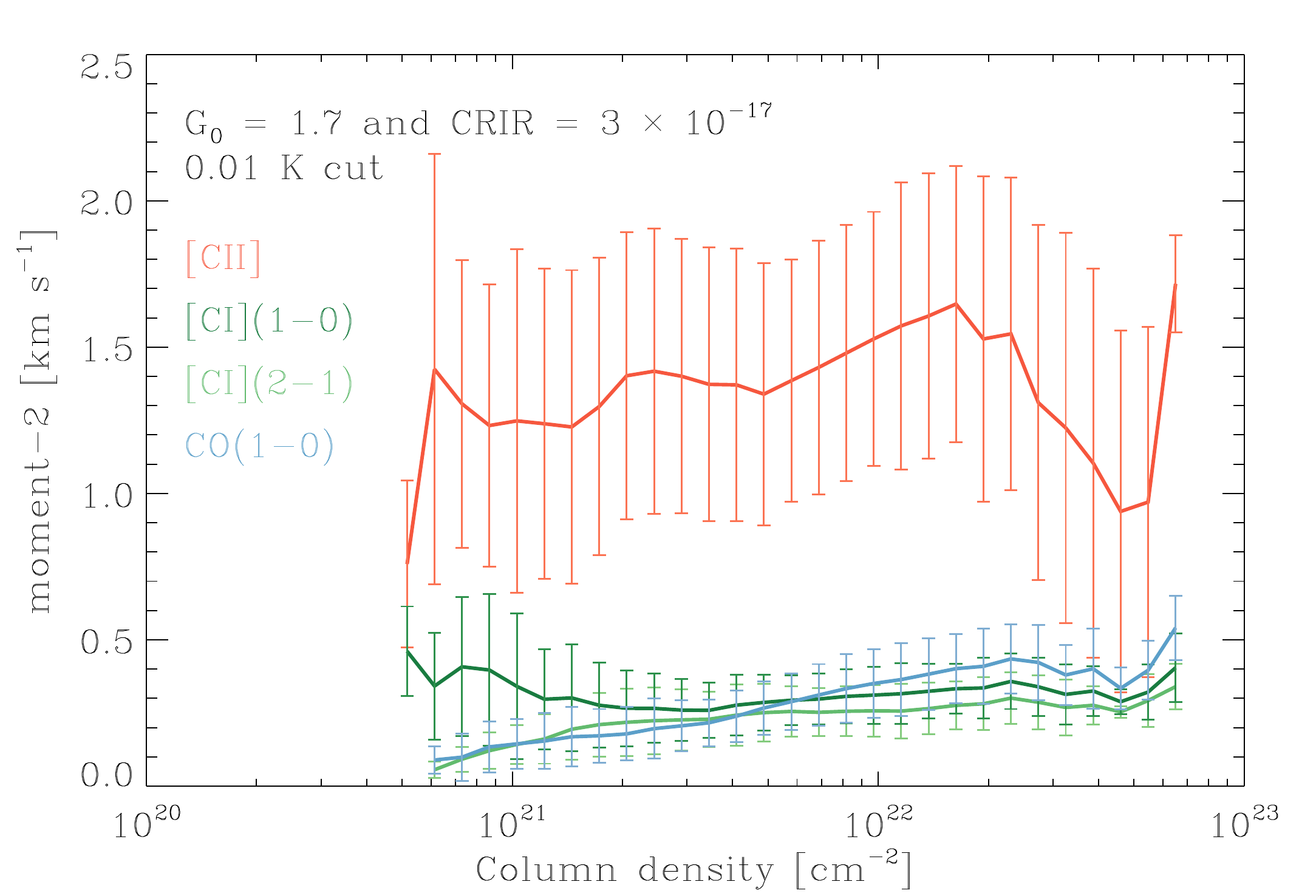}
\includegraphics[width=3.2in]{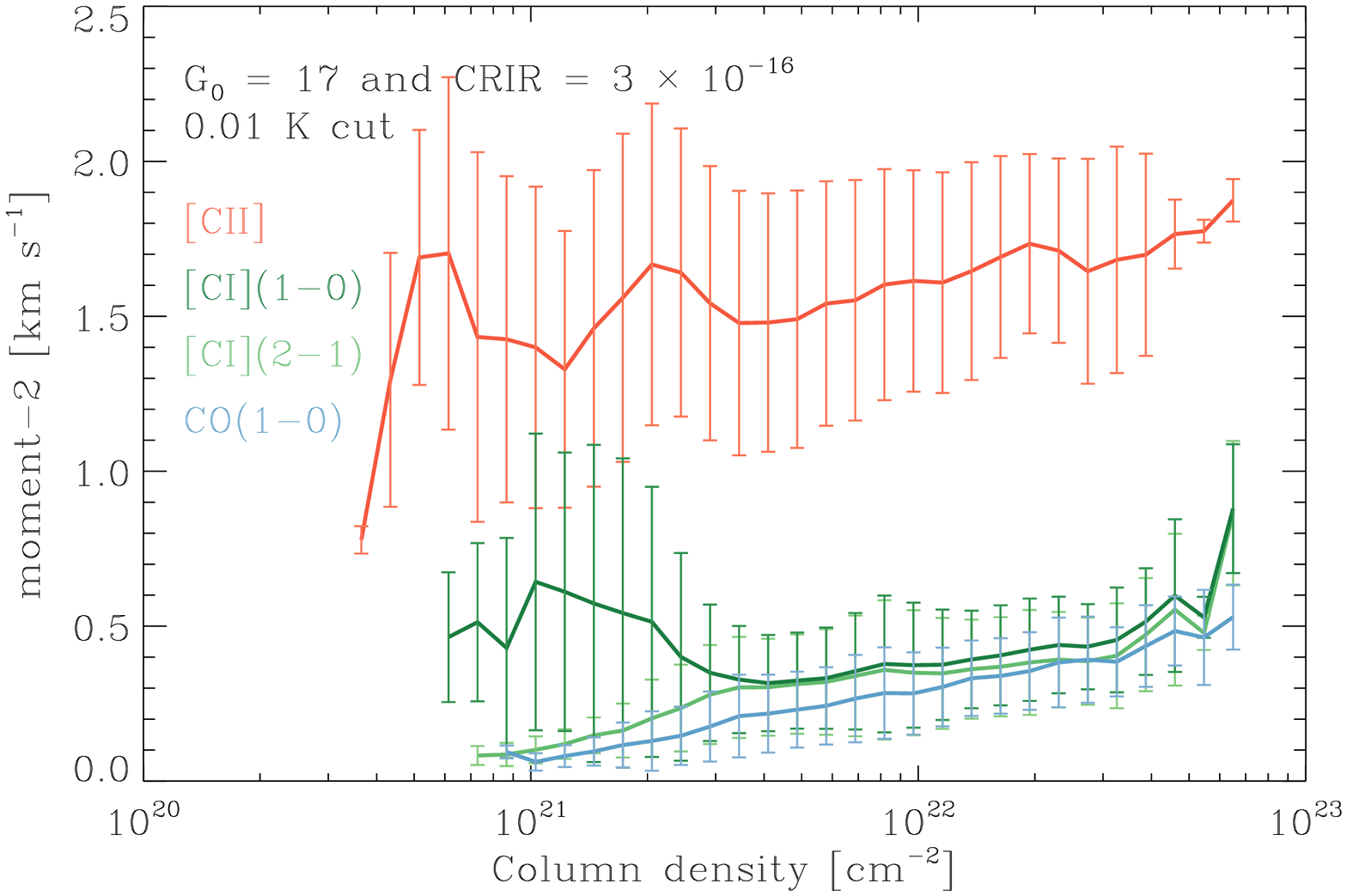}
\caption{{\em Top}: mean value of the second moment -- as given by Equation \ref{eq:mom-2} -- for each of our tracers, plotted as a function of the total column density, for the run with the fiducial values of G$_{0}$ and the CRIR.
The ``error bars'' indicate the standard deviation at each column density. The colour-coding of the different lines is the same as in the previous figures. {\em Bottom}: as in the top panel, but for the run with ten times larger values for G$_{0}$ and the CRIR. }
\label{fig:mom2_col}
\end{figure}

\section{Velocity information}
\label{sec:vels}
In our discussion so far, we have seen that the [CII] traces a different regime to the [CI] and CO, probing gas that is of lower density, slightly higher temperature and of mainly atomic, rather than molecular, composition.  Since our simulations involve an initially well-ordered flow, it is worth investigating whether these differences also arise in the velocity information that is contained in the line emission. 

We start by looking at the information contained in the second moment, defined via
\begin{equation}
\label{eq:mom-2}
\sigma_{\rm mom-2}(i, j) = \left[ \frac{\sum_{v = v_{\rm min}}^{v_{\rm max}}T(i, j, v) (v - v_{\rm cent}(i, j))^2 }{\sum_{v = v_{\rm min}}^{v_{\rm max}}T(i, j, v)} \right]^{1/2}
\end{equation}
where $i$ and $j$ denote pixels in $y$ and $z$ directions in the image, $v$ is the velocity of a channel, $T$ is the brightness temperature of the emission in the voxel $(i, j, v)$, and $v_{\rm cent}(i, j)$ is the centroid velocity, given by,
\begin{equation}
\label{eq:mom-1}
v_{\rm cent}(i, j) =  \frac{\sum_{v = v_{\rm min}}^{v_{\rm max}}T(i, j, v) v }{ \sum_{v = v_{\rm min}}^{v_{\rm max}}T(i, j, v)}.
\end{equation}
The second moment is essentially an emission-weighted version of the velocity dispersion along the line-of-sight, while the centroid velocity (the first moment) is the emission-weighted mean velocity.  Note that we only include voxels in our emission cubes that have emission above 0.01 K. These measures are more useful than line-fitting in cases where there are several velocity components in the image, which as we will see shortly, is very much the case in our colliding clouds. 

The mean and standard deviation of $\sigma_{\rm mom-2}$ as a function of column density in the image are given in Figure \ref{fig:mom2_col} for each of our lines.  The first thing we see is that $\sigma_{\rm mom-2}$ for the [CII] line is generally higher than that from both [CI] and CO lines. In principle this can mean one of two things: either the gas traced by [CII] is hotter, and so we are seeing the effects of thermal line-broadening in the second moment, or the [CII] line is originating from gas that has multiple velocity components or a higher velocity dispersion. However, we know from our previous analysis that most of the [CII] emission is being produced in gas which is colder than 100~K. The thermal velocity of the C$^{+}$ ions at this temperature is approximately $0.3 \, {\rm km \, s^{-1}}$, and hence thermal broadening cannot be responsible for the differences in $\sigma_{\rm mom-2}$ between [CII] and CO or [CI] that we see at some column densities. These differences must therefore be due primarily to differences in the bulk motions of the gas traced by the different forms of emission. Further support for this conclusion comes from the fact that we see only relatively minor differences in $\sigma_{\rm mom-2}$ in the runs with different ISRF strengths and CRIRs, despite the substantial differences in the temperature structure of these two runs.

We also see that $\sigma_{\rm mom-2}$ for CO and [CI], rises with increasing column density. This is interesting, since it goes in the opposite direction to the ``{\em transition to coherence}'' picture that is often claimed for prestellar cores in molecular clouds (see e.g.\ \citealt{good98}, \citealt{kirk07}, J.~E.~Pineda~et~al.~\citeyear{jaime10}), although this transition is generally seen in high density tracers such as N$_{2}$H$^{+}$ or NH$_{3}$, rather than [CI] or $^{12}$CO.  The fact that the bulk of the CO emission is coming from gas around a few 1000 cm$^{-3}$ -- around the critical density -- means that the line-width of  any given parcel of emitting gas will be only 0.2 km s$^{-1}$, the sound speed at around 10 K. That we see $\sigma_{\rm mom-2} > 0.2 \rm \, km \, s^{-1}$ implies that we are seeing non-thermal contributions from multiple parcels of gas along the line-of-sight. There are two reasons why $\sigma_{\rm mom-2}$ would then  increase with increasing column density.  The first is that higher column density regions can be the result of more violent compression, and so what we are seeing is the residual velocity components that have survived the shock. Only head-on collisions in uniform flows would result in a post-shock velocity of zero, and even then, only if the gas remains thermally stable; oblique shocks will always have left over shear, the amount of which will scale with the with strength of the shock, which also increases the density of the gas.  The second reason that we see a higher $\sigma_{\rm mom-2}$ with increasing column density could simply be due to the additional influence of gravity as we go to progressively denser structures.
 
It is also interesting that the behaviour seen in Figure \ref{fig:mom2_col} is the opposite from what we see in Figure \ref{fig:X10_images}, where the {\em density-weighted} velocity dispersion in the regions of high-column is markedly lower than that found in the regions of high density. Given that we expect line emission to scale linearly with density in the LTE regime and with the density squared for sub-thermal excitation, this is clearly a worry for our observational interpretation of the gas velocities in molecular clouds.  Indeed, it suggests that a more careful decomposition of the line emission data is required to make sense of the kinematics,  such as the technique presented in \citet{Henshaw2016}.
 
\begin{figure*}
\centerline{ \includegraphics[width=7.3in]{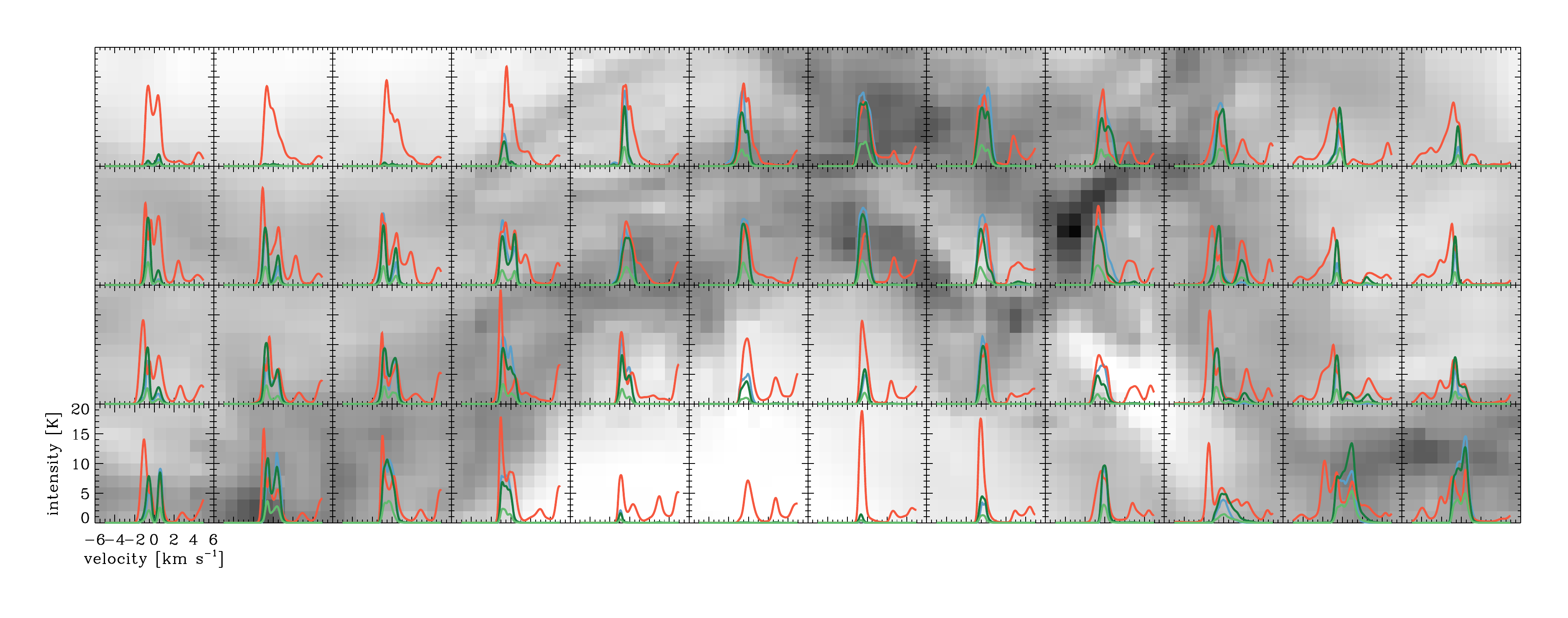} }
\caption{The background grayscale shows a zoom of the column density around the bottom ridge that appears in Figure 1 for our simulation with G$_0 = 17$ and a CRIR $ = 3 \times 10^{-16} \rm\, s^{-1}$.  Superimposed on the column density image are spectra for each of the emission lines examined in our study, averaged over the region shown by the axes, which cover a region of $10 \times 10 $ pixels, or 0.24~pc $\times$ 0.24~pc.  In keeping with our previous colour palette, reddish-orange is the [CII] line, dark and light green are, respectively,  the (1-0) and (2-1) transitions of [CI], and CO (1-0) is in blue. Note that for clarity the intensity of the [CII] line has been multiplied by a factor 20, and those of the [CI] lines have been multiplied by 2.}
\label{fig:spectra} 
\end{figure*}

\begin{figure*}
\centerline{
	\includegraphics[width=2.3in]{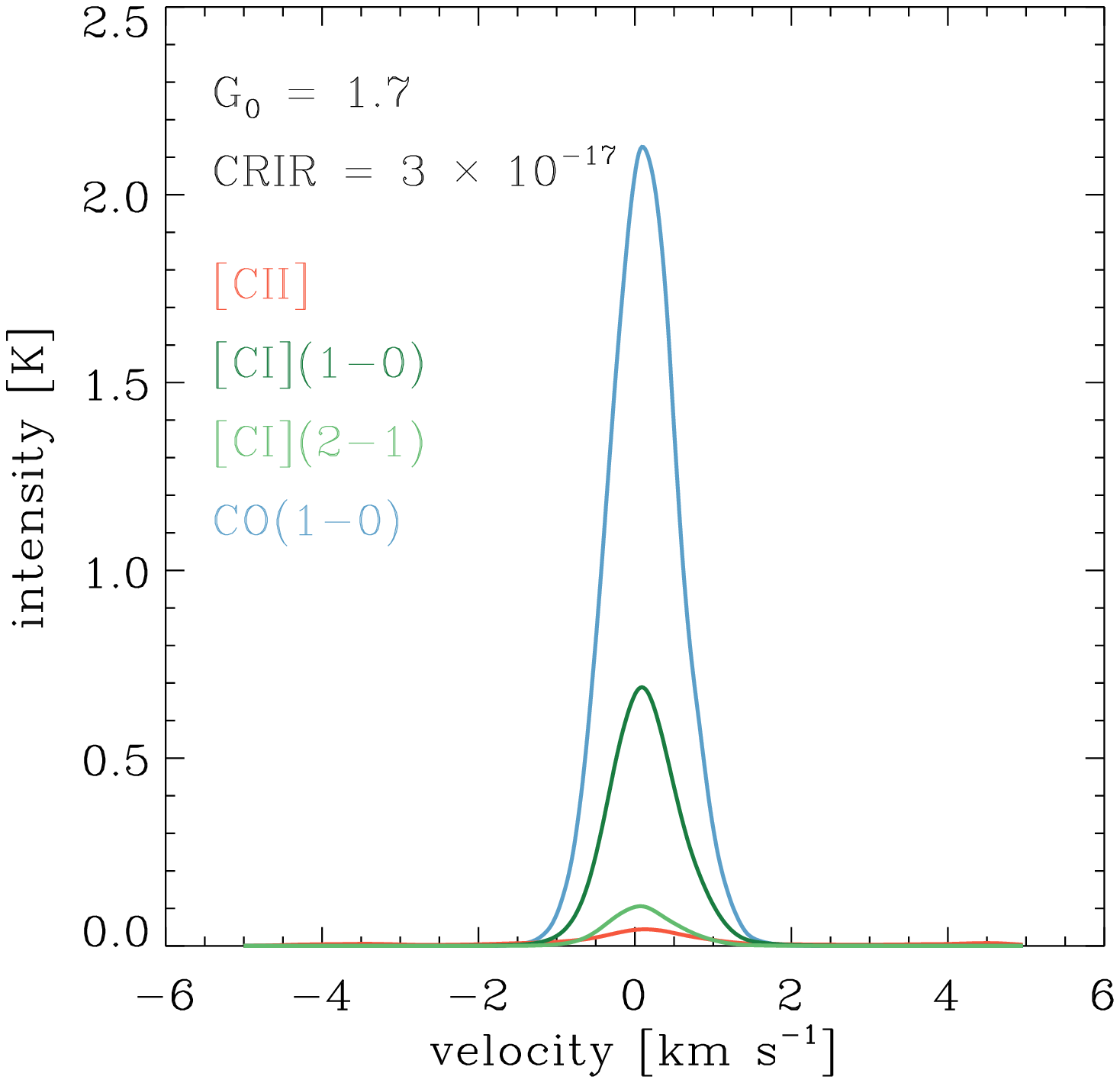}
	\includegraphics[width=2.3in]{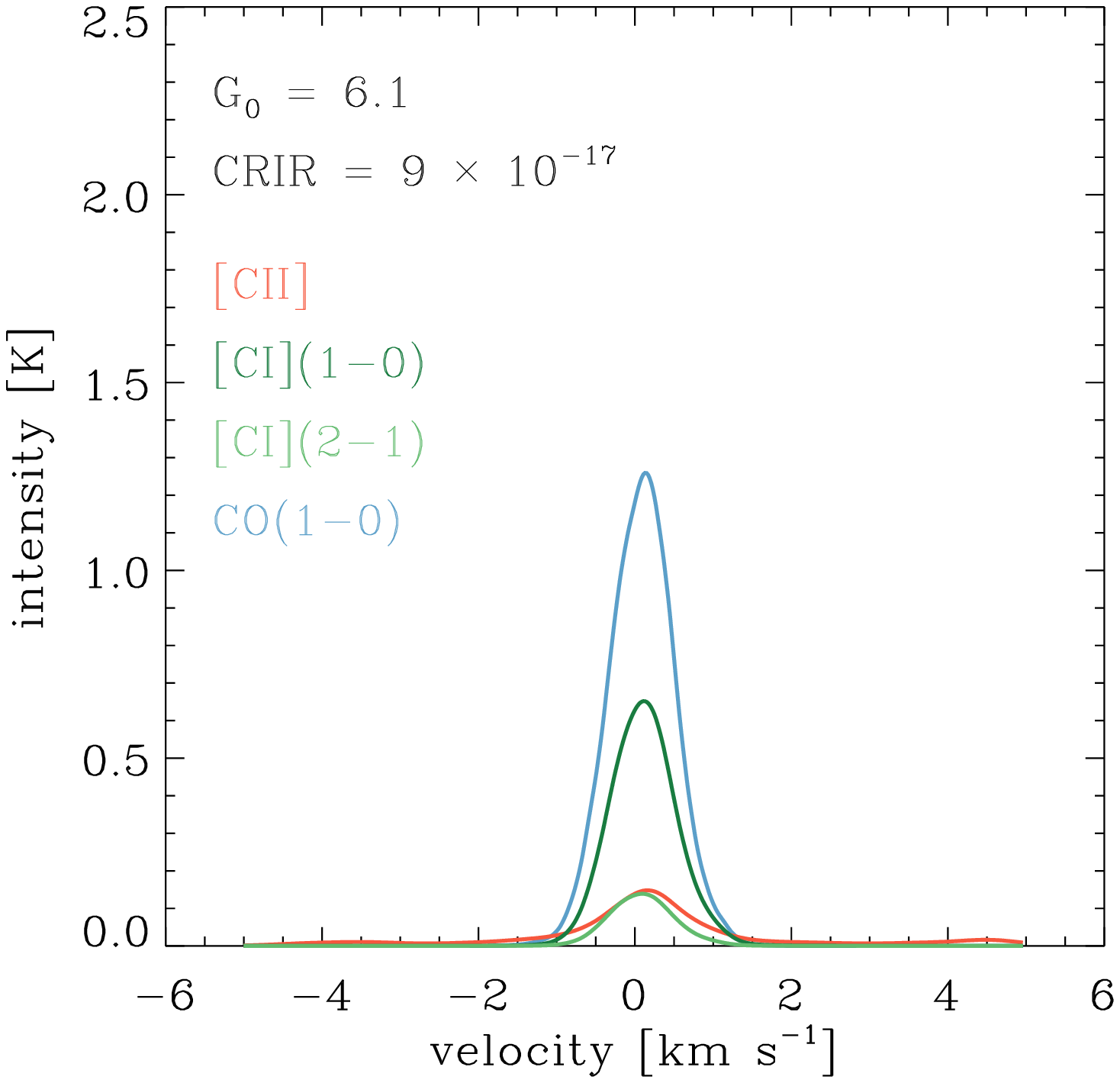}
	\includegraphics[width=2.3in]{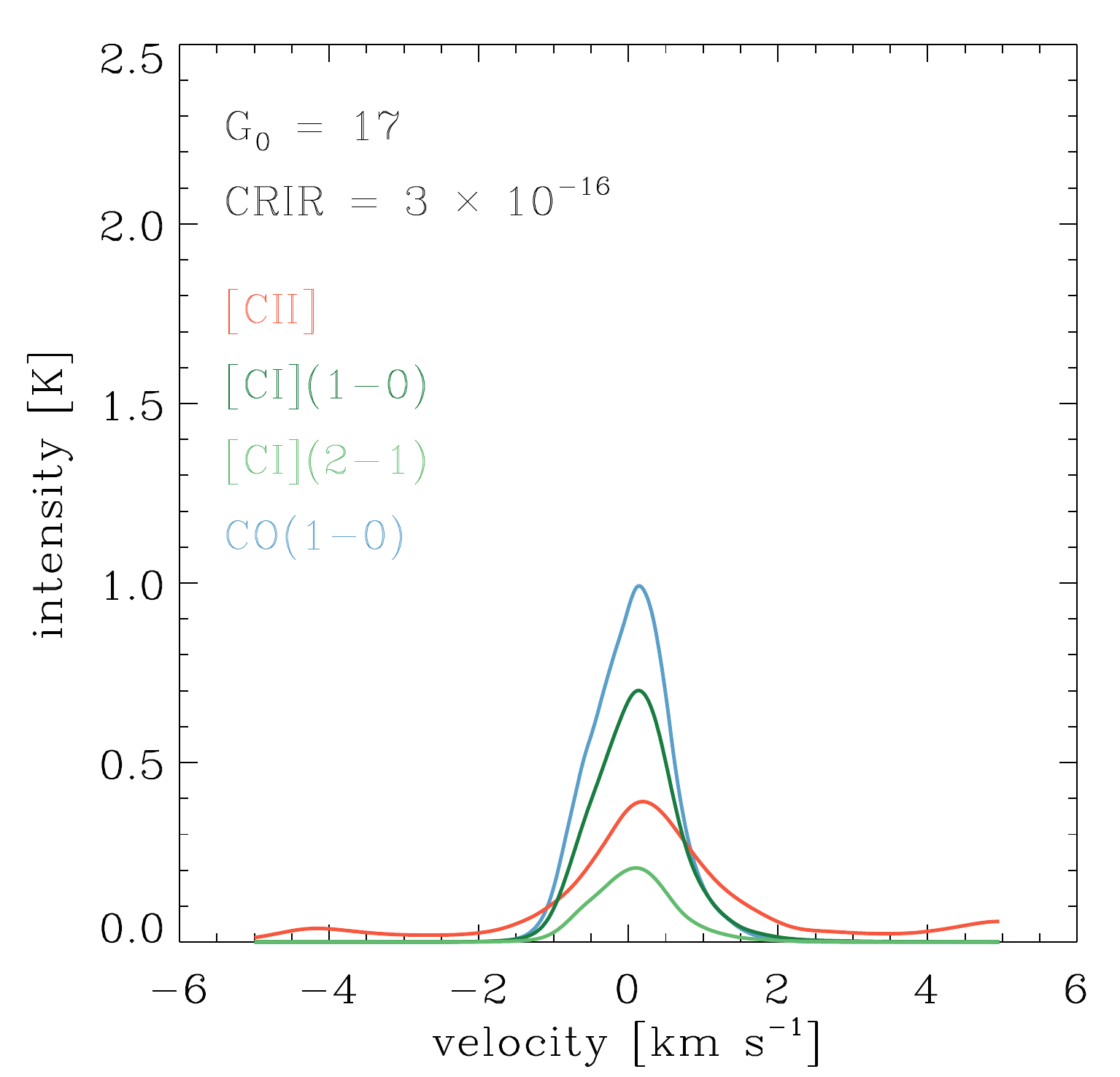}
}
\caption{Mean spectra from the maps shown in Figure \ref{fig:mom0}.}
\label{fig:cloudspec}
\end{figure*}

\begin{figure*}
\centerline{ 
\includegraphics[width=3.2in]{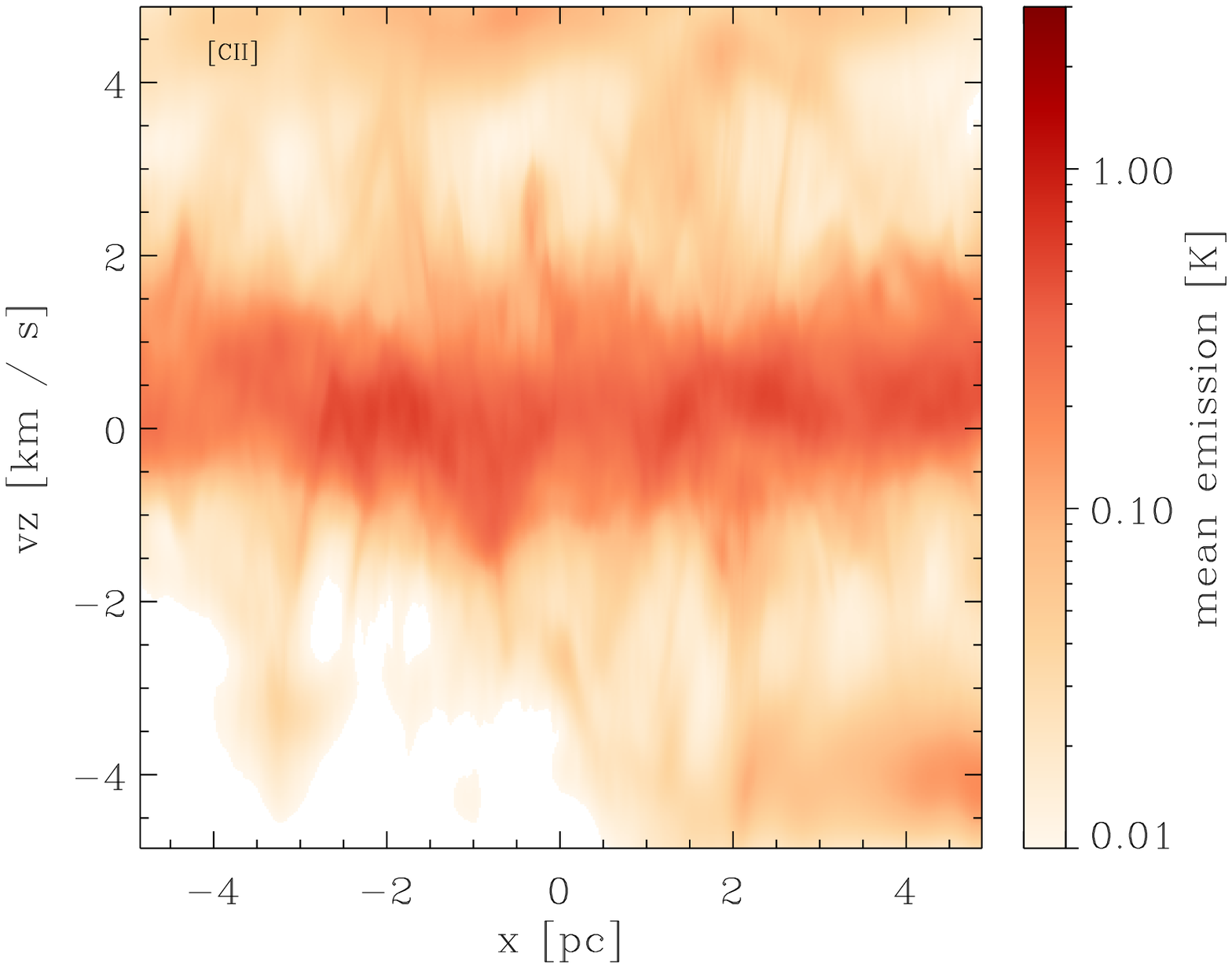}
\includegraphics[width=3.2in]{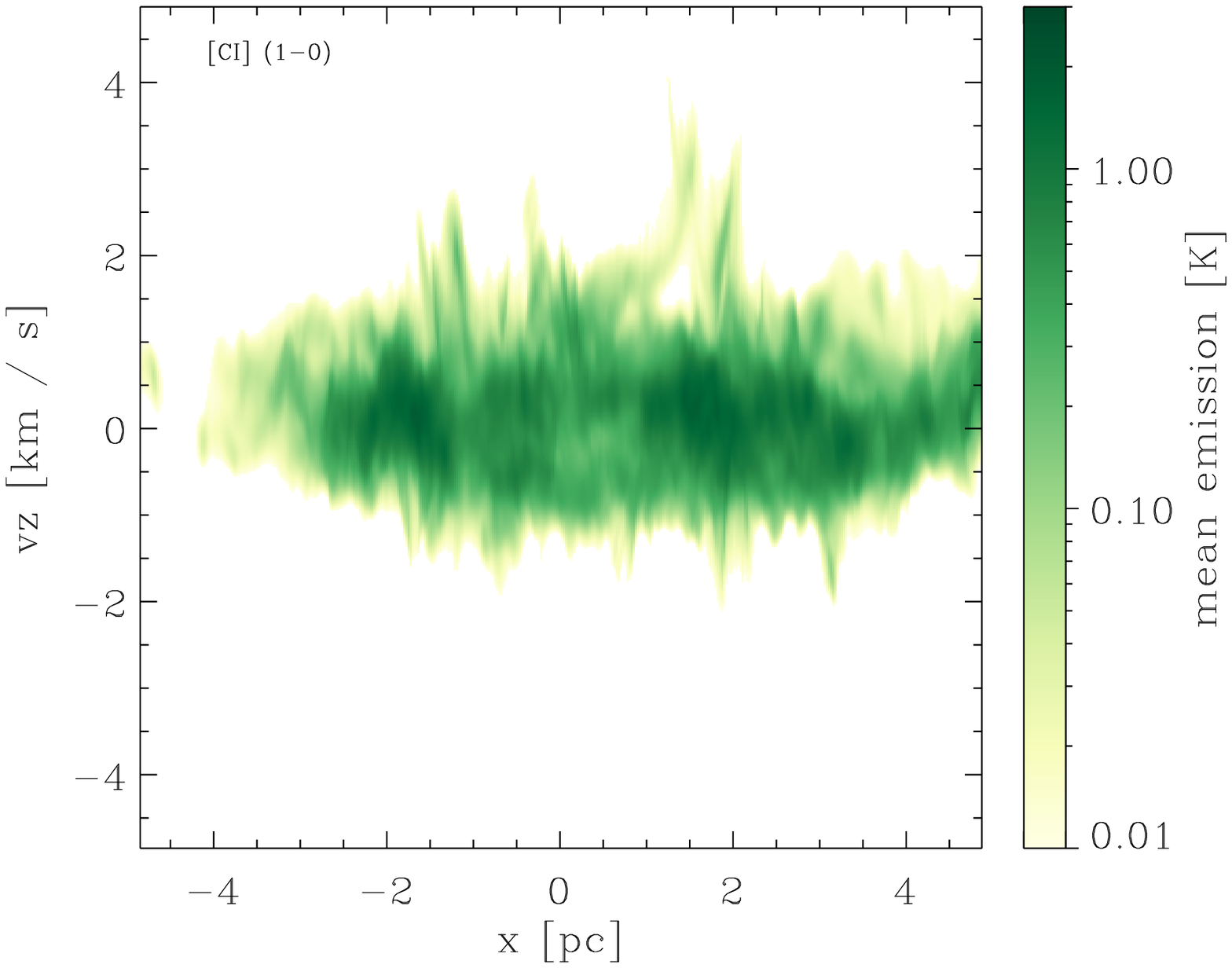}
}
\centerline{ 
\includegraphics[width=3.2in]{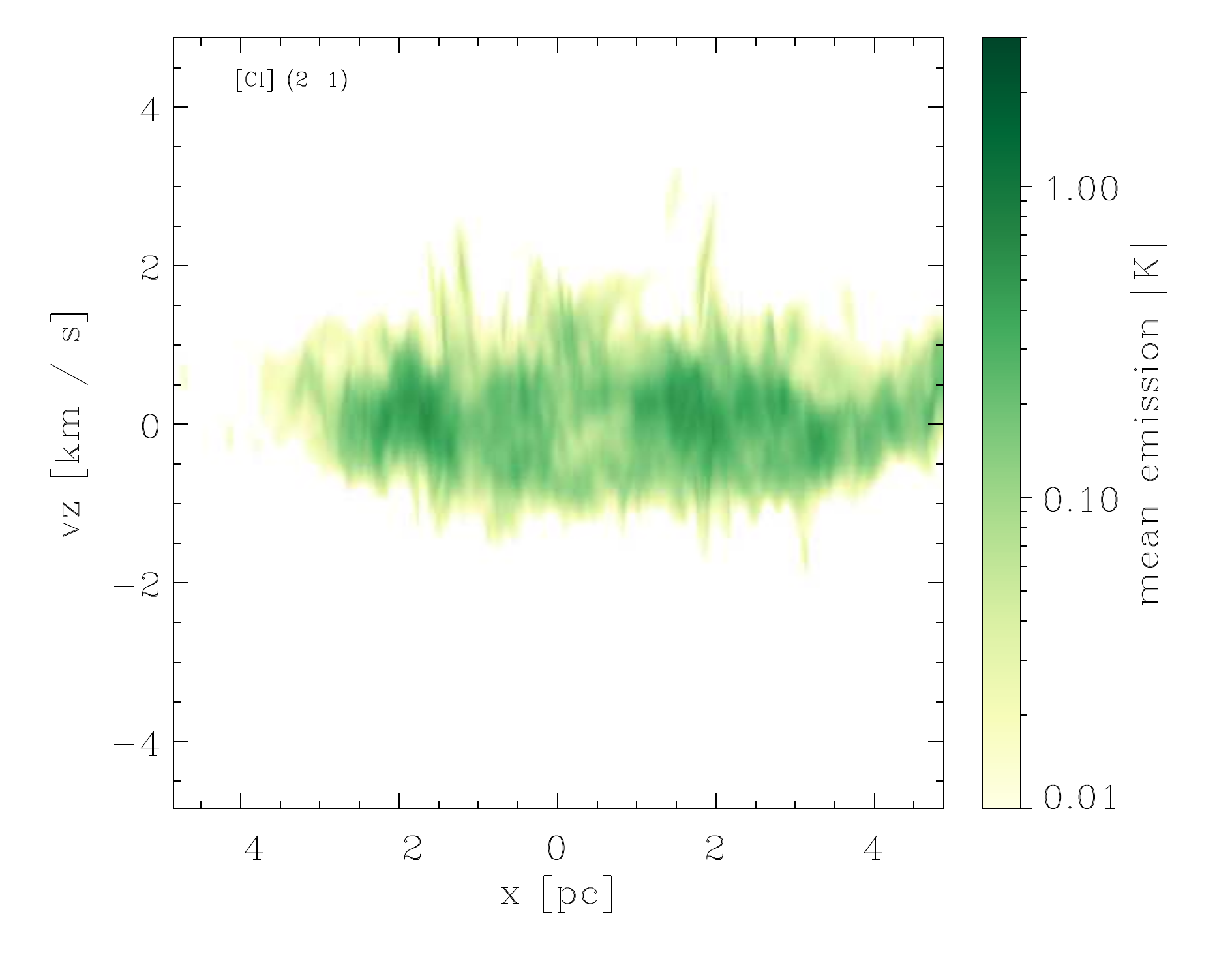}
\includegraphics[width=3.2in]{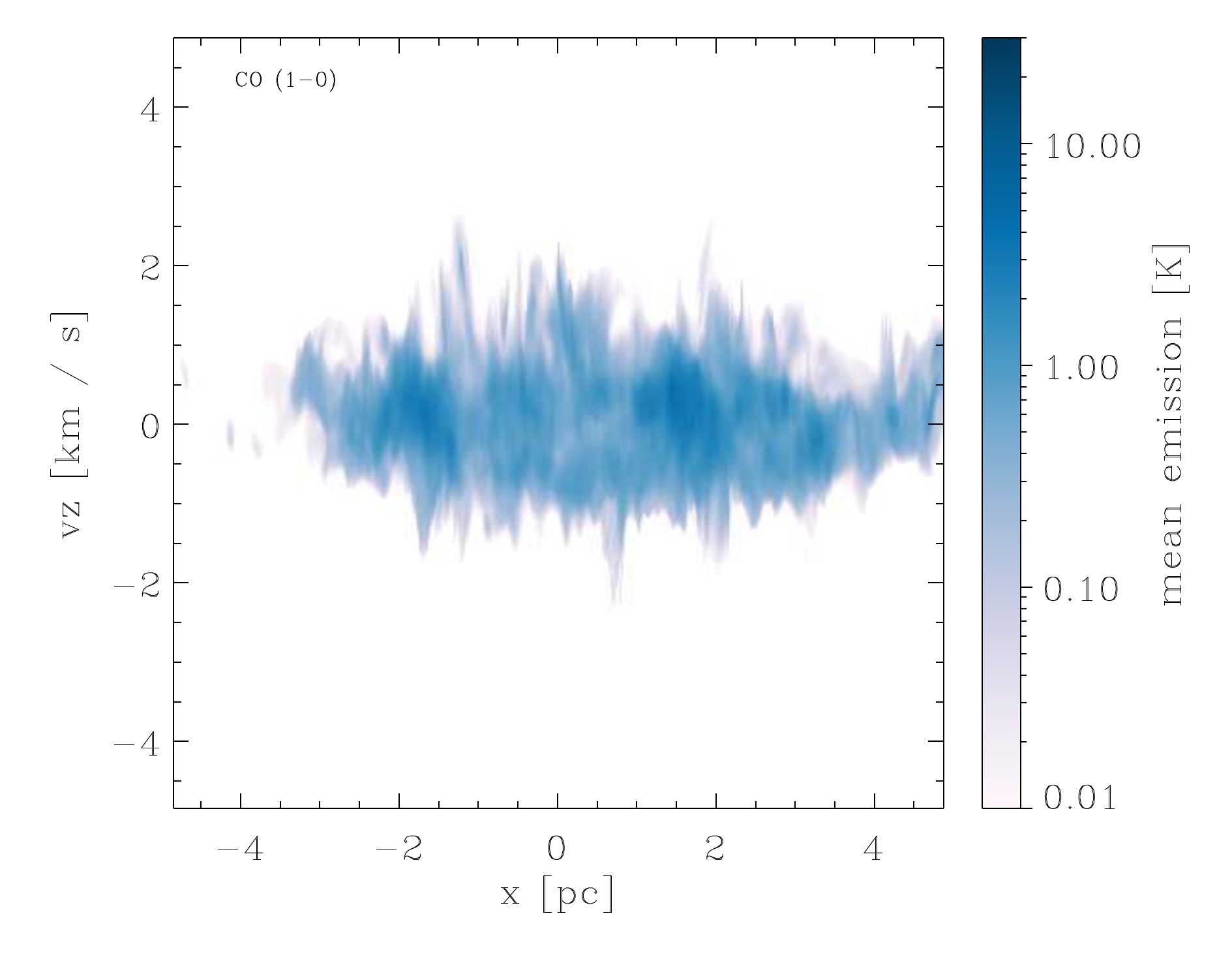}
}
\caption{Position-velocity diagrams for the simulation with G$_0 = 17$ and a CRIR $ = 3 \times 10^{-16} \rm\, s^{-1}$. Note that the colour scale for the CO (1-0) emission is stretched over a larger range of mean brightness temperatures than for the other lines.}
\label{fig:posvelX10}
\end{figure*}

Although the [CII] emission traces higher values of $\sigma_{\rm mom-2}$ than CO and [CI], it is still unclear from our analysis as to whether these are coherent features or the result of multiple velocity components along the line-of-sight. In addition, our idea to explain the variation in $\sigma_{\rm mom-2}$ with column density for the CO emission requires multiple components. We can get more information by looking at the spectra themselves. In Figure \ref{fig:spectra} we show a zoomed-in region from our simulation with the high ISRF and CRIR. The greyscale in the background is the column density from the top panel in Figure \ref{fig:X10_images}. The panels in the Figure also show the averaged spectra for our four lines in the area covered by the panel. For clarity, we have multiplied the intensity of the [CII] line by a factor of 20, and that of the two [CI] lines by a factor of 2.  

The first obvious feature is that the [CII] line is significantly broader than the other lines, and that it is always made up of multiple velocity components -- i.e. there are several peaks in [CII] emission along each line-of-sight. However, we also see that the majority of these peaks are quite narrow, similar to the widths of the features in the CO and [CI] lines. This confirms our analysis above, demonstrating that the difference in $\sigma_{\rm mom-2}$ between [CII] and the other lines is driven by differences in the gas velocities, rather than the temperature of the gas. These peaks thus represent structures at different velocities along each line of sight. The fact that we can see them in the [CII] emission, but not in the other lines is important: it demonstrates that [CII] is able to trace the structure of the (mainly) atomic gas as the molecular clouds are being assembled. 

We see that the [CII] components with large velocity offsets from their CO or [CI] counterparts are generally quite faint. Given that our initial setup involved $10$ cm$^{-3}$ clouds with velocities of $\pm 3.75$ km s$^{-1}$, one would conclude that [CII] is not doing a good job of picking up the original flows that created the cloud. Indeed, what we have seen in our discussion above is that very little emission is coming from gas with densities close to $n = 10 \, \rm cm^{-3}$, the density of our initial conditions.  Therefore, although [CII] is able to trace this inflowing material, successfully detecting the very faint emission coming from this gas will be very challenging given current observational capabilities, and mapping it over an extended region may require a prohibitively large amount of observing time. Current observational capabilities are instead better suited to studying the emission from the denser ($n \sim 100 \, \rm cm^{-3}$) CO-poor gas immediately surrounding the CO-rich portions of the cloud.

There are several other features of these spectra that are worth noting. First, we see that there is a huge variation in the line shapes, total intensities and relative intensities, even in this small region of around 3 pc $\times$ 1 pc, similar to the variation observed by {\em Herschel} in the GOT-C$^+$ survey \citep{pineda13}.  We also note that the spectra vary rapidly from panel to panel (which are around 0.24 pc on a side),  especially when we have sharp variations in the background column density.   Perhaps more important, however, is the fact that [CII] is always present, even at positions in $ppv$ space that have strong CO and [CI] emission. However, we do see that there is typically an offset in velocity between the [CII] peaks and those of the other lines, again suggesting that inside molecular regions the bulk of the [CII] is still tracing slightly different gas. 

In Figure \ref{fig:cloudspec} we show the spectra for the entire cloud -- that is, the spectra averaged over the area shown in Figure \ref{fig:mom0} -- and include now all three simulations. In contrast to Figure~\ref{fig:spectra}, these averaged spectra are shown without rescaling. \citet{bisbas17b} showed that the peaks in the [CII] and CO spectra were offset in the cloud-cloud collisions that they studied.  Although this is the case for our spectra in Figure \ref{fig:spectra} which looks at the small scales within the cloud exposed to the high ISRF and CRIR, it is not the case when we study the full cloud spectra shown in Figure \ref{fig:cloudspec}; while the full-cloud spectra are generally asymmetric, the peaks in the spectra are actually well aligned.  However, we do see that the [CII] linewidths are broader than those for [CI] and CO. 

\citet{bisbas17b} also examined the position velocity (p-v) diagrams of cloud-cloud collisions --  a common technique for looking at cloud-cloud collision signatures, which take the form of ``bridges'' in the p-v diagram between loci of emission (c.f. \citealt{DC2011, Haworth2015}) -- and found clear signatures of the collision in their analysis.  In Figure \ref{fig:posvelX10} we show the p-v diagram for our high ISFR and CRIR simulation. In all tracers, we see that most of the emission is coming from the central ridge that sits around zero velocity, and this is probably why the full-cloud spectra have their peaks at the same velocity. However we also see vertical striations in all the plots, showing that there are large velocity dispersions at many locations in the cloud, and that these features exist in {\em all} our tracers.  When we look at the p-v diagram for the [CII] emission,  see that these striations bridge the central, bright, zero velocity ridge and the two fainter ridges at $\pm 4 \rm \, km\,s^{-1}$. These higher velocity ridges correspond to original atomic clouds, although they have been shifted from (and spread around) their original $\pm 3.75  \rm \, km\,s^{-1}$ by the turbulence that we impose in the initial conditions.

Figure \ref{fig:posvelX10} once again demonstrates that our original atomic clouds are very difficult to detect, since the brightness temperature of the emission from them is typically $\lesssim 0.1$ K, even for this case of a high ISRF.  In our runs with weaker ISRFs, this extended emission is even fainter. In the study by \citet{bisbas17b}, the [CII] emission in the bridging features was of similar strength to our study, even though they adopted a smaller radiation field strength, G$_0 \sim 4$. However the clouds in their simulation start at a number density of 100 cm$^{-1}$, 10 times higher than our clouds, and this higher density helps to compensate for the weaker radiation field. Taken together, our two studies suggest that bridging emission from [CII] is generally weak in regions where the ISRF is low.

\section{Discussion}
\label{sec:discuss}
The {\em Herschel} GOT C$^{+}$ survey of [CII] emission in the Galactic plane (J.~L.~Pineda~et~al.~\citeyear{pineda10}, \citealt{pineda13}) provides us with what is currently the most comprehensive set of velocity-resolved observations of [CII] in the Milky Way. \citet{pineda13} used this data set, together with existing surveys of the Galactic plane at other wavelengths, to study the association between [CII] and HI, CO and star formation. One of the key results of this study was the finding that in the Milky Way, only around half of the observed [CII] emission is directly associated with star-forming clouds, with most of this emission coming from dense photodissociation regions (PDRs) and a small fraction from HII regions. The other half of the observed [CII] emission arises in gas clouds illuminated by relatively weak radiation fields (G$_{0} = 1$--10). \citet{pineda13} attempt to distinguish between emission coming from cold atomic gas clouds and emission coming from clouds dominated by CO-dark H$_{2}$, concluding that roughly 60\% of the [CII] emission that is not directly associated with star formation comes from H$_{2}$-dominated clouds, with the remainder coming from clouds dominated by atomic gas. 

In comparison, in this study we find that a large fraction of the [CII] emission in our synthetic images comes from gas dominated by atomic hydrogen, and only a small fraction comes from gas dominated by H$_{2}$. What is the reason for this discrepancy? Part of the reason may be the assumption made in \citet{pineda13} that all of the cold HI has a spin temperature of 100~K, which leads them to conclude that the correction that must be made for optical depth effects when deriving the HI column density is small. In practice, we find that much of the HI in our clouds is colder than this (see Figures~\ref{fig:rhotemp} \& \ref{fig:rhoxcx}), and so \citet{pineda13} are arguably underestimating the optical depth correction required, and hence underestimating the amount of [CII] emission that can be associated with the HI. Future studies of the optical depth of cold HI in the Galactic plane with e.g.\ the THOR survey \citep{beuther16} will help to clarify this. Nevertheless, it seems unlikely that this can explain all of the difference between our simulations and the observational results. 

A more promising explanation for the difference is that it is a consequence of the evolutionary state of the clouds. The gas clouds sampled in the GOT C$^{+}$ survey are not selected to be in any particular evolutionary state and hence may have a wide range of ages. On the other hand, the clouds simulated in our study are young, since we halt the simulation at the point at which we expect star formation to begin. Given the relatively long formation timescale for H$_{2}$ at densities below $n \sim 100 \: {\rm cm^{-3}}$, it is likely that the gas producing the bulk of the [CII] emission in our simulations has not yet reached chemical equilibrium, and that older clouds may produce more of their [CII] in H$_{2}$-dominated gas and less in H-dominated gas.  Resolved studies of individual clouds do indicate that the bulk of [CII] emission often is not associated with molecular material \citep[e.g.][]{Beuther2014,Perez2015}, but further such studies are needed to determine whether an evolutionary trend exists.

Support for this picture comes from two other recent studies of [CII] emission from simulated molecular clouds. \citet{franeck18} produce synthetic [CII] emission maps of one of the clouds modelled in the SILCC-Zoom project of \citet{ds17}. As the name suggests, this project involves the carrying out of ``zoom-in'' simulations that follow the formation and growth of a small set of clouds formed in the large-scale simulations of the Galactic plane performed as part of the SILCC project\footnote{https://hera.ph1.uni-koeln.de/$\sim$silcc/} (see e.g.\ \citealt{walch15}, \citealt{girichidis16} for more details). Those zoom-in simulations reach a very high spatial resolution ($\Delta x < 0.1$~pc), comparable to that in our study, but are carried out with a different hydrodynamical code (the FLASH AMR code; \citealt{flash}) and with far less idealized initial conditions. They therefore provide a useful point of comparison to our own study. In common with our study, \citet{ds17} end their simulations shortly before the onset of star formation in the clouds, and so the [CII] maps produced by \citet{franeck18} once again correspond to the emission we expect from a relatively young cloud. Interestingly, \citet{franeck18} also find that most of the [CII] emission in their simulations comes from regions dominated by atomic hydrogen, with less than 20\% coming from H$_{2}$-dominated gas. This is in line with what we would expect if this is a common feature of dynamically young clouds.

The amount of [CII] emission arising from H-dominated and H$_{2}$-dominated clouds in Milky Way-like conditions was also examined by \citet{gs16}. They post-processed the high-resolution galactic-scale simulations of \citet{smith14} to produce synthetic [CII] and [OI] emission maps, making the important simplifying assumption that the emission was optically thin. They then examined the distribution of this emission as a function of variables such as the atomic hydrogen fraction or the gas temperature. As the \citet{smith14} simulations did not include star formation or stellar feedback, the results of the \citet{gs16} can be compared directly to the \citet{pineda13} results for quiescent clouds. \citet{gs16} find that roughly half of the [CII] emission in their model comes from H$_{2}$-dominated gas, with the remainder coming from H-dominated gas. This is broadly comparable to the 60/40 split found by \citet{pineda13}, but contrasts strongly with our results and those of \citet{franeck18}. However, this is just what we would expect if the strength of the association between [CII] and HI varies with the age of the clouds. The synthetic emission maps produced by \citet{gs16} contain clouds with a wide range of dynamical ages, as do the GOT C$^{+}$ observations, and so we should not expect their results to match those of simulations that purely study young clouds.

Finally, it is worth asking how robust our results are, given that we have used a simplified chemical network to track the evolution of carbon in the clouds. The NL99 network that adopt here has been shown to give similar results to the more advanced network of \cite{glover10} for solar neighbourhood values of the ISRF strength and CRIR \citep{gc12b}. However, \citet{gong17} demonstrated that it does not perform as well under conditions with higher UV fields and CRIRs, and suggested some improvements, which we will refer to as GOW17. In particular, they found that the GOW17 modifications yield CO abundances that better match the results of a more sophisticated PDR code at low A$_{\rm V}$ than those from NL99. Naturally, this could implications for our study.  In our preliminary testing of the GOW17 network, we actually find results that are very similar to those we present here with the NL99 network, and so our conclusions here would not change. The main differences are that i) both [CI] lines trace very slightly lower densities than CO(1-0), but only by a factor of $\sim$ 0.6, and ii) that there is a very small amount of low density emission in both CO and [CI], although some of this is also likely attributed to the ``channel blending'' that we discussed earlier. However, our conclusions -- i.e. that [CII] traces a different regime to both [CI] and CO (which trace very similar conditions) -- remain unchanged.  We plan to explore the GOW17 network more fully in future work.

\section{Conclusions}
\label{sec:conc}
We present a study of the [CII], [CI] and CO emission associated with the collision of two initially atomic clouds of $10^4 \rm \, M_{\odot}$, and explore how the emission varies as we vary the ISRF and CRIR.  Our clouds start with an initial density of 10 cm$^{-3}$, similar to the value found in the CNM \citep{wolfire03}. 
The clouds are each given a velocity of 3.75 $\rm km\, s^{-1}$ towards the other cloud. Given that the sound speed in our gas is between $0.6 \, \rm km\, s^{-1}$ and $1.7 \, \rm km\, s^{-1}$, depending on the strength of the ISRF, this ensures that the collision in our initial setup has a Mach number above 2. The ISRF and CRIR  are scaled together from the solar neighbourhood values of $G_0 = 1.7$ and $3 \times 10^{-17} \rm s^{-1}$, respectively, to 3 and ten times these values. Our simulations are performed with {\sc Arepo} \citep{springel10}, and use a detailed model of the ISM chemistry and thermodynamics \citep[see e.g.][]{gc12,smith14}.

The simulations are stopped once the first collapsing core appears in each case. As this point, we use the RADMC-3D radiative transfer code to produce synthetic emission maps of the [CII] and [CI] fine structure lines and the $J = 1 \rightarrow 0$ transition of CO. We compare the $ppv$ cubes of emission to $ppv$ cubes of the density, temperature, and molecular fraction in our clouds in order to determine what physical conditions our lines are tracing, and also examine the kinematics revealed by the different tracers. All of these comparisons are made looking along the axis of the collision (i.e.\ perpendicular to the shock plane). Our main results are as follows:
\begin{itemize}
\item All our simulations resulted in the formation of a network of dense molecular (i.e.\ H$_2$-rich) structures, with around 50\% of the initial hydrogen mass being converted to molecular form by the point at which we stop the simulations. The amount of molecular gas formed has little dependence on the strength of the ISRF or the CRIR.

\item The gas makes the transition from H to H$_2$ at around a number density of a few 100 cm$^{-3}$, irrespective of the CRIR and the strength of the ISRF.  However the carbon chemistry changes occur at much higher densities. For the solar neighbourhood ISRF and CRIR, C$^+$ is still the dominant form of carbon at a number density of 1000 cm$^{-3}$, and CO does not become dominant until a density of nearly $10^4$  cm$^{-3}$. When the ISRF and CRIR are 10 times higher, these transitions are shifted up in density by a factor of $\sim 5$. At no point is the neutral form of carbon the dominant form. 

\item We find that majority of the [CII] emission comes from gas with number densities $\sim 100$ cm$^{-3}$, which is still predominantly atomic in nature. At higher densities the gas is too cold to excite the line, and at lower densities the emissivities are too small to be readily detectable with current observing platforms.  However, we note that gas at this density has not yet reached chemical equilibrium in our simulations, and it is therefore plausible that if we were to examine much older clouds, we would find a much larger fraction of the [CII] emission coming from H$_{2}$-dominated gas.

\item The [CI] and CO emission are very similar and trace gas that is predominantly molecular in nature.  Most of the emission from neutral carbon and CO comes from gas with number density 500 -- 1000$ \, \rm cm^{-3}$ and temperature $< 30$~K. 

\item For our simulation with solar neighbourhood values of the ISRF and CRIR, the peak brightness temperature of the [CII] emission is only 0.33 K, and the highest integrated intensity is 0.23 K km s$^{-1}$. This is only marginally detectable with upGREAT on SOFIA. At 3 times the solar neighbourhood ISRF and CRIR, we get a peak brightness temperature of 0.71 K and a maximum integrated intensity of 0.64 K km s$^{-1}$, while at 10 times the solar neighbourhood ISRF and CRIR, these increase to 2.66~K and 2.55~K~km~s$^{-1}$, respectively.

\item We find that the velocity dispersion of the [CII] emission is larger than that of  the CO and [CI] emission. [CII] traces additional, widely-spaced velocity components in the spectra of the molecular cloud that are not seen in the other tracers. We also see evidence of ``bridging'' features in the position-velocity diagrams that show that [CII] emission is coherently extended beyond the CO and [CI] emission, similar to those seen in the study of  \citet{bisbas17b}. 
Although this shows that [CII] is able to trace the flows that form the molecular gas, the emission is very faint in the bridging features, and would currently be difficult to detect in most cases.

\item Although the [CI] emission traces gas at slightly lower column densities than CO, we do not find it to be a better tracer of the collision than the CO. We do not find any significant [CI] emission coming from the atomic (or weakly molecular) gas that constitutes the original colliding clouds.

\end{itemize}

In summary, 
we find that although [CII] is a good tracer of the atomic clouds just before the molecular transition, it is currently very difficult to observe this phase if the clouds have low densities, particularly for ISRF strengths close to the standard solar neighbourhood value. On the other hand, [CII] emission from colder, denser atomic gas associated with the already assembled portion of the cloud should be much easier to observe with current facilities. Our results also emphasise the importance of the local radiation field strength for determining the strength of the [CII] emission, with the important implication that clouds forming in regions with elevated radiation fields will be much easier to trace in [CII] than clouds in quiescent regions.

\section*{Acknowledgments}
PCC and ADC acknowledge support from the Science and Technology Facilities Council (under grant ST/N00706/1). We also acknowledge StarFormMapper, a project that has received funding from the European Union's Horizon 2020 Research and Innovation Programme, under grant agreement no. 687528. SCOG acknowledges financial support from the Deutsche Forschungsgemeinschaft  via SFB 881, ``The Milky Way System'' (sub-projects B1, B2 and B8) and SPP 1573, ``Physics of the Interstellar Medium''. SCOG also acknowledges support from the European Research Council under the European Community's Seventh Framework Programme (FP7/2007-2013) via the ERC Advanced Grant STARLIGHT (project number 339177). SER acknowledges support from the European Union's Horizon 2020 research and innovation programme under the Marie Sk{\l}odowska-Curie grant agreement \# 706390.

\appendix
\section{Details of the chemical model}
\label{app:chem}
The chemical model used in the simulations presented in this paper is an updated version of the NL99 network from \citet{gc12}. This itself is a combination of two separate chemical networks, the simplified carbon and oxygen network developed by \citet{nl99} and the hydrogen chemistry network developed by \citet{gm07a,gm07b}. A full list of the reactions contained in our current version of the NL99 network is given in Table~\ref{tab:chem}, along with references to the sources from which we took the rate coefficients for each reaction. 

The first part of the table (reactions 1--24) lists the reactions taken from \citet{nl99}. The artificial chemical species CH$_{\rm x}$ and OH$_{\rm x}$ involved in some of these reactions represent in an approximate fashion small carbon-carrying molecular ions and radicals (e.g.\ CH, CH$_{2}$, CH$^{+}$) and oxygen-carrying molecular ions and radicals (e.g.\ OH, OH$^{+}$, etc.), respectively. The artificial species M represents several different low ionization potential metals (e.g.\ Na, Mg) that are the dominant gas-phase charge carriers in dense and well-shielded gas. 

We have made one minor modification to this part of the network compared to the version in \citet{nl99}. The original version of the network includes the following pseudo-reaction instead of reactions 1 and 2:
\begin{equation}
{\rm H_{2} + cr + H_{2}} \rightarrow {\rm H_{3}^{+} + e^{-} + H}.
\end{equation}
This pseudo-reaction represents the fact that in the conditions that they study, where all of the hydrogen is molecular, almost all of the H$_{2}^{+}$ ions produced by cosmic ray ionization of H$_{2}$ will then react with other H$_{2}$ molecules to form H$_{3}^{+}$. We cannot make the same assumption, as we want our network to be useable in conditions where not all of the hydrogen has yet been incorporated into H$_{2}$. Therefore, we include reactions 1 and 2 as separate reactions, rather than using \citeauthor{nl99}'s pseudo-reaction, and we also account for H$_{2}^{+}$ destruction by charge transfer with atomic hydrogen (reaction 25), which is an important process when the atomic hydrogen fraction is large.

The other minor difference between the reactions listed in the first part of Table~\ref{tab:chem} and the ones in the original \citet{nl99} paper involves reaction 18, the destruction of H$_{3}^{+}$ by charge transfer with M, which represents a combination of different low ionization potential metals (Mg, Na, etc.; see the discussion in \citealt{nl99} for more details). In their paper, \citet{nl99} give this reaction as
\begin{equation}
{\rm H_{3}^{+} + M \rightarrow M^{+} + e^{-} + H_{2}},
\end{equation}
but this is evidently a typo, since as written neither charge nor the number of hydrogen atoms is conserved. We list the reaction below using the corrected form
\begin{equation}
{\rm H_{3}^{+} + M \rightarrow M^{+} + H + H_{2}}.
\end{equation}

\begin{table}
\caption{List of reactions included in our chemical model \label{tab:chem}}
\begin{tabular}{rll}
\hline
No.\ & Reaction & Reference \\
\hline
1 & ${\rm H_{2} + cr \rightarrow H_{2}^{+} + e^{-}}$ & See text \\
2 & ${\rm H_{2}^{+} + H_{2} \rightarrow H_{3}^{+} + H}$ & SLD98 \\
3 & ${\rm He + cr \rightarrow He^{+} + e^{-}}$ & See text \\
4 & ${\rm H_{3}^{+} + C \rightarrow CH_{x} + H_{2}}$ & NL99 \\
5 & ${\rm H_{3}^{+} + O \rightarrow OH_{x} + H_{2}}$ & NL99 \\
6 & ${\rm H_{3}^{+} + CO \rightarrow HCO^{+} + H_{2}}$ & NL99 \\
7 & ${\rm He^{+} + H_{2} \rightarrow He + H + H^{+}}$ & B84 \\
8 & ${\rm He^{+} + CO \rightarrow C^{+} + O + He}$ & P89 \\
9 & ${\rm C^{+} + H_{2} \rightarrow CH_{x} + H}$ & NL99 \\
10 & ${\rm C^{+} + OH_{x} \rightarrow HCO^{+}}$ & NL99 \\
11 & ${\rm O + CH_{x} \rightarrow CO + H}$ & NL99 \\
12 & ${\rm C + OH_{x} \rightarrow CO + H}$ & NL99 \\
13 & ${\rm He^{+} + e^{-} \rightarrow He + \gamma}$ & HS98 \\
14 & ${\rm H_{3}^{+} + e^{-} \rightarrow H_{2} + H}$ & MAC04 \\
15 & ${\rm C^{+} + e^{-} \rightarrow C + \gamma}$ & NP97 \\
16 & ${\rm HCO^{+} + e^{-} \rightarrow CO + H}$ & GEP05 \\
17 & ${\rm M^{+} + e^{-} \rightarrow M + \gamma}$ & NL99 \\
18 & ${\rm H_{3}^{+} + M \rightarrow M^{+} + H + H_{2}}$ & NL99 \\ 
19 & ${\rm C + \gamma \rightarrow C^{+} + e^{-}}$ & NL99 \\
20 & ${\rm CH_{x} + \gamma \rightarrow C + H}$ & NL99 \\
21 & ${\rm CO + \gamma \rightarrow C + O}$ & V09 \\
22 & ${\rm OH_{x} + \gamma \rightarrow O + H}$ & NL99 \\
23 & ${\rm M + \gamma \rightarrow M^{+} + e^{-}}$ & NL99 \\
24 & ${\rm HCO^{+} + \gamma \rightarrow CO + H}$ & NL99 \\
\hline
25 & ${\rm H_{2}^{+} + H \rightarrow H_{2} + H^{+}}$ & KAH79 \\
26 & ${\rm H_{3}^{+} + e^{-} \rightarrow H + H + H}$ & MAC04 \\
27 & ${\rm H + e^{-} \rightarrow H^{+} + e^{-} + e^{-}}$ & A97 \\
28 & ${\rm H^{+} + e^{-} \rightarrow H + \gamma}$ & FER92 \\
29 & ${\rm H^{+} + e^{-}(s) \rightarrow H + \gamma}$ & WD01 \\
30 & ${\rm He^{+} + H_{2} \rightarrow He + H_{2}^{+}}$ & B84 \\
31 & ${\rm H_{2} + H \rightarrow H + H + H}$ & MS86 \\
32 & ${\rm H_{2} + H_{2} \rightarrow H + H + H_{2}}$ & MKM98 \\
33 & ${\rm H_{2} + e^{-} \rightarrow H + H + e^{-}}$ & TT02 \\
34 & ${\rm H + H(s) \rightarrow H_{2}}$ & HM89 \\ 
35 & ${\rm C + H_{2} \rightarrow CH_{x} + \gamma}$ & PH80 \\
36 & ${\rm HCO^{+} + e^{-} \rightarrow CH_{x} + O}$ & GEP05 \\
37 & ${\rm H_{2} + \gamma \rightarrow H + H}$ & DB96 \\
38 & ${\rm H + cr \rightarrow H^{+} + e^{-}}$ & See text \\
39 & ${\rm C + cr \rightarrow C^{+} + e^{-}}$ & See text \\
40 & ${\rm C + \gamma_{cr} \rightarrow C^{+} + e^{-}}$ & See text \\
41 & ${\rm CO + \gamma_{cr} \rightarrow C + O}$ & See text \\
\hline
\end{tabular}
\\ The reactions listed above the line are the same as those in the original \citet{nl99} chemical model (with two minor alterations, discussed in the text), although in some cases the reaction rate coefficients we use differ from those in their model. The reactions below the line were not included in the original \citet{nl99} model. `cr' represents a cosmic ray, $\gamma$ a photon from the ISRF, and $\gamma_{\rm cr}$ a UV photon produced by excitation of H or H$_{2}$ by the high energy secondary electrons produced by the cosmic ray ionization of H, He or H$_{2}$. `(s)' indicates that the species in question is adsorbed on the surface of a dust grain. The meaning of the chemical symbols CH$_{\rm x}$, OH$_{\rm x}$ and M is discussed in the text. \\
{\bf References}: A97 -- \citet{a97}; B84 -- \citet{b84}; DB96 -- \citet{db96}; FER92 -- \citet{fer92}; GEP05 -- \citet{gep05}; HM89 -- \citet{hm89}; HS98 -- \citet{hs98}; KAH79 -- \citet{kah79}; MAC04 -- \citet{mac04}; MKM98 -- \citet{mkm98}; MS86 -- \citet{ms86}; NL99 -- \citet{nl99}; NP97 -- \citet{np97}; P89 -- \citet{p89}; PH80 -- \citet{ph80};  SLD98 -- \citet{sld98}; TT02 -- \citet{tt02}; V09 -- \citet{visser09};  WD01 -- \citet{wd01}
\end{table}

In the second half of Table~\ref{tab:chem}, we list the reactions that we have added to the original \citet{nl99} set to make the combined NL99 network. Many of these dealing with the chemistry of H$^{+}$, H and H$_{2}$ come from the \citet{gm07a,gm07b} chemical network, but we have also supplemented these with a number of other reactions from various sources. Of particular note is the inclusion of cosmic-ray induced photodissociation of CO (reaction 41), which can in some circumstances dominate the destruction of CO in gas which is well-shielded from the ISRF (see e.g.\ \citealt{mac18}). 

The rate coefficients for the reactions in the portion of the network based on \citet{nl99} are largely taken from that work. In the cases where they are not, this is either because more accurate values based on experiment or theory have subsequently become available (e.g.\ reactions 14, 16, 21), or because our adopted rate coefficients are valid over a wider range of temperatures than those given in \citet{nl99}. The rate coefficients for the other portion of the network are drawn from a variety of sources, as summarized in Table~\ref{tab:chem}.

One set of reactions deserves further comment, those involving cosmic ray ionization or cosmic-ray induced photodissociation. In the case of the cosmic ray ionization reactions (numbers 1, 3, 38, and 39), we first specify the cosmic ray ionization for neutral hydrogen (reaction 38) as an input parameter to the simulation and then set the rates for the other processes by using scaling factors derived from the rates given in \citet{rate12}. In the case of the two cosmic-ray induced photodissociation reactions (numbers 40 and 41), we follow the same procedure for reaction 40, but for reaction 41 we adopt the scaling factor given in \citet{mht96}, which is a fit to calculations by \citet{g89}. 

Finally, we note that we do not account for the freeze-out of CO onto dust grains in our current chemical model. This process can have a profound impact on the gas-phase CO abundance in highly-shielded dense gas with a low dust temperature. However, as previous studies have already shown \citep[see e.g.][]{gold01,gc16}, it has a minimal impact on the $^{12}$CO emission observed at a great distance from the shielded gas, as the regions where freeze-out is significant are generally also highly optically thick in the $^{12}$CO lines. We therefore would not expect the inclusion of this process to significantly change our results. 

\section{Details of the thermal model}
As is common in simulations of the interstellar medium that do not adopt an isothermal equation of state, we model the thermal evolution of the gas using an operator split approach. The effects of adiabatic expansion and contraction of the gas, as well as viscous dissipation in shocks, are accounted for as part of the standard hydrodynamical treatment, as discussed in detail in \citet{springel10}. However, during each timestep we also account for the impact of radiative and chemical heating and cooling on the internal energy density of the gas $\epsilon$ by solving an ordinary differential equation (ODE) of the form:
\begin{equation}
\frac{{\rm d}\epsilon}{{\rm d}t} = - \Lambda(\rho, \epsilon, x_{\rm H_{2}}, x_{\rm H^{+}}, ...).  \label{eq:energy}
\end{equation}
Here, $\Lambda$ is the net cooling rate per unit volume due to both radiative and chemical processes. Processes that result in heating (e.g.\ photoelectric emission from dust grains) are included by treating them as negative cooling. As $\Lambda$ depends not only on the mass density and internal energy density of the gas, but also on its chemical composition, we solve Equation~\ref{eq:energy} in parallel with the chemical rate equations using the DVODE solver \citep{dvode}.

The full list of processes included in the current version of the thermal model was recently published in \citet{mac18}, and so on the grounds of brevity we do not include it here. Instead, we simply note that the only significant difference between the thermal model presented in that paper and the one used for the simulations presented here is the absence of X-ray Coulomb heating in our simulations. This process is not included simply because we are considering a situation in which there is not a significant X-ray background.

As well as solving for the gas temperature, we also solve for the dust temperature on the fly in our simulations, as this is important for determining the H$_{2}$ formation rate on dust grains and the rate at which collisions transfer thermal energy between the gas and the dust. The details of our dust temperature calculation are the same as those described in Appendix~A of \citet{gc12}. As dust cooling is very efficient, we assume that the dust temperature is always at its equilibrium value, which can be found by balancing the effects of dust heating due to the ISRF, dust cooling due to its own thermal radiation, and collisional energy transfer between the gas and the dust (which heats the dust if $T_{\rm gas} > T_{\rm dust}$, or cools it if $T_{\rm dust} > T_{\rm gas}$). Shielding of the ISRF, which lowers the effectiveness of dust heating in regions with high $A_{\rm V}$, is accounted for using the {\sc treecol} algorithm. As \citet{cgk12} demonstrate, the resulting dust temperatures agree well with those computed using a more sophisticated Monte Carlo treatment.

\end{document}